\documentclass[a4paper]{amsart}
\usepackage{latexsym}
\usepackage{amsmath}
\usepackage{amssymb}
\usepackage{amsfonts}

\newcounter{smallarabics}
\newenvironment{arabicenumerate}
{\begin{list}{{\normalfont\textrm{(\arabic{smallarabics})}}}
  {\usecounter{smallarabics}\setlength{\itemindent}{0cm}
   \setlength{\leftmargin}{5ex}\setlength{\labelwidth}{4ex}
   \setlength{\topsep}{0.75\parsep}\setlength{\partopsep}{0ex}
   \setlength{\itemsep}{0ex}}}
{\end{list}}

\newcounter{smallroman}

\newcommand{\ben}{\begin{arabicenumerate}}  
\newcommand{\een}{\end{arabicenumerate}}

\def\init{\setcounter{equation}{0}}

\newtheorem{theoreme}{Theorem }[section]
\newtheorem{proposition}[theoreme]{Proposition}
\newtheorem{lemma}[theoreme]{Lemma}
\newtheorem{definition}[theoreme]{Definition}

\newtheorem{remark}[theoreme]{Remark}
\newtheorem{example}[theoreme]{Example}
\newcommand{\beq}{\begin{equation}}
\newcommand{\eeq}{\end{equation}}

\newcommand{\bex}{\begin{example}}
\newcommand{\eex}{\end{example}}
\def\bel{\begin{lemma}}
\def\eel{\end{lemma}}
\def\bet{\begin{theoreme}}
\def\eet{\end{theoreme}}
\def\bed{\begin{definition}}
\def\eed{\end{definition}}
\def\ber{\begin{remark}}
\def\eer{\end{remark}}

\def\rr{{\mathbb R}}

\def\cc{{\mathbb C}}
\def\nn{{\mathbb N}}

\def\slim{{\rm s-}\lim}
\def\cl{{\rm cl}}
\def\H{{\rm H}}

\def\al{\alpha}
\def\i{{\rm i}}
\def\Span{{\rm Span}}
\def\qed{$\Box$\medskip}
\def\Sp{{\mathcal Sp}}
\def\D{\mathcal D}
\def\Ran{\mathrm{Ran}}
\def\bbbone{{\mathchoice {\rm 1\mskip-4mu l} {\rm 1\mskip-4mu l}
{\rm 1\mskip-4.5mu l} {\rm 1\mskip-5mu l}}}
\def\one{\bbbone}
\def\cH{{\mathcal H}}
\def\C{Q}
\def\R{{\langle R\rangle}}

\def \p{ \partial}

\def\12{\frac{1}{2}}

\def\xt{\frac{x}{t}}
\def\supp{{\rm supp}}

\def\e{{\mathrm e}}

\def\bep{\begin{proposition}}
\def\eep{\end{proposition}}

\newcommand{\ti}{\tilde}

\newcommand{\De}{\Delta}

\newcommand{\mcL}{\mathcal{L}}

\newcommand{\pa}{\partial}

\newcommand{\Om}{\Omega}

\newcommand{\hil}{\mathcal{H}}
\newcommand{\om}{\omega}
\newcommand{\mfa}{\mathfrak{A}}
\newcommand{\mco}{\mathcal{O}}

\newcommand{\eps}{\varepsilon}
\newcommand{\fr}[2]{\frac{#1}{#2}}

\newcommand{\real}{\mathbb{R}}

\newcommand{\la}{\lambda}
\newcommand{\ov}{\overline}

\newcommand{\non}{\nonumber}

\newcommand{\beqa}{\begin{eqnarray}}
\newcommand{\eeqa}{\end{eqnarray}}
\newcommand{\hx}{\tilde x}
\def\tom{\tilde{\omega}}

\newcommand{\setbar}{\ : \ }
\newcommand{\BB}{B_1,B_2}
\newcommand{\BBc}{B_2,B_1}
\newcommand{\wh}{\widehat}
\newcommand{\timeso}{\overset{\mathrm{out}}{\times}}

\def\rx{x}
\def\xt{\frac{x}{t}}

\def\rr{\real}
\def\cH{\hil}
\def\bbbone{{\mathchoice {\rm 1\mskip-4mu l} {\rm 1\mskip-4mu l}
{\rm 1\mskip-4.5mu l} {\rm 1\mskip-5mu l}}}
\def\one{\bbbone}
\def\cS{\mathcal{S}}
\def\supp{{\rm supp}}

\def\p{\partial}
\def\nn{{\mathbb{N}}}
\def\12{\frac{1}{2}}
\def\Dom{{\rm Dom}\ }

\def\coinf{C_{0}^{\infty}}
\def\cc{{\mathbb{C}}}
\def\tom{\tilde{\omega}}

\usepackage{verbatim}

\DeclareFontFamily{U}{mathx}{\hyphenchar\font45}
\DeclareFontShape{U}{mathx}{m}{n}{
      <5> <6> <7> <8> <9> <10>
      <10.95> <12> <14.4> <17.28> <20.74> <24.88>
      mathx10
      }{}
\DeclareSymbolFont{mathx}{U}{mathx}{m}{n}
\DeclareFontSubstitution{U}{mathx}{m}{n}
\DeclareMathAccent{\widecheck}{0}{mathx}{"71}
\DeclareMathAccent{\wideparen}{0}{mathx}{"75}

\newlength{\dinwidth}
\newlength{\dinmargin}
\begin{document}
\title[Asymptotic completeness of two-particle scattering]{Towards asymptotic completeness of two-particle scattering in local relativistic QFT}
\author{Wojciech Dybalski}
\address{Zentrum Mathematik, Technische Universit\"at M\"unchen,
D-85747 Garching Germany}
\email{dybalski@ma.tum.de}
\author{Christian G\'erard}
\address{D\'epartement de Math\'ematiques, Universit\'e de Paris XI, 91405 Orsay Cedex France}
\email{christian.gerard@math.u-psud.fr}
\begin{abstract} 
We consider the problem of existence of asymptotic observables   
in local relativistic theories of massive particles. Let $\ti p_1$ and $\ti p_2$ 
be two energy-momentum vectors
of a massive particle and let $\De$ be a small neighbourhood of $\ti p_1+\ti p_2$. We construct asymptotic observables (two-particle Araki-Haag detectors), sensitive to neutral particles of energy-momenta
in small neighbourhoods of $\ti p_1$ and $\ti p_2$. We show that these asymptotic observables  exist, as strong limits of 
their approximating sequences, on all
physical states from the spectral subspace of  $\De$.  Moreover, the linear span of the ranges of all such asymptotic observables coincides with the subspace of two-particle Haag-Ruelle scattering states with total energy-momenta  in $\De$. The result holds under very general conditions which are satisfied, for example, in $\la \phi^4_2$.  The proof of convergence relies on a  variant of the phase-space propagation estimate of Graf.
\end{abstract}
\maketitle

\section{Introduction}\label{Intro}
\setcounter{equation}{0}

The question of  a complete particle interpretation of quantum theories is of fundamental importance
for our understanding of physics. The solution of this problem in non-relativistic quantum mechanics, 
obtained in
 \cite{En78, SiSo87,Gr90, De93}
 for a large class of physically relevant Hamiltonians, requires the convergence
of   suitably chosen time-dependent families of observables. The existence of these limits, called {\em asymptotic observables},
relies on the method of {\em propagation estimates} \cite{SiSo87,Gr90}, which is a refined variant of the Cook method. 
This technique  was later adapted to non-relativistic QFT in \cite{DG99} which initiated a systematic study of the problem of asymptotic completeness in this context \cite{DG00, FGS02, FGS04, DM12}. In the present work we implement the method of propagation estimates in  local relativistic
quantum field theories of massive particles. We obtain the existence of certain asymptotic observables 
which can be interpreted as two-particle detectors. Our results, stated in Theorems~\ref{Main-theorem} and 
\ref{Weak-asymptotic-completeness} below, hold in any massive theory 
satisfying   the Haag-Kastler axioms, for example in $\la\phi^4_2$. Our work sheds a new light  on the problem
of asymptotic completeness in such theories, which is widely open to date.

The problem of existence of asymptotic observables in the framework of algebraic quantum field theory (cf. Subsection~\ref{general-assumptions}) 
was first studied in the seminal work of Araki and Haag \cite{AH67} and later by Enss in \cite{En75}. These authors considered families of  observables 
of the form
\beqa
C_t:=\int h\left(\xt\right)C(t,x)dx=\e^{\i tH} \int h\left(\xt\right)C(x)dx\,\e^{-\i tH}, \label{C-t-definition}
\eeqa
where $C$ denotes a suitable (almost local) observable, $C(t,x)$ its translation in space-time by $(t,x)\in \real^{1+d}$, $H$ is the full Hamiltonian
of the relativistic theory and $h\in C_0^{\infty}(\real^{d})$. 
They were  able to show that products of such observables
\beqa
\C_{n,t}= C_{1,t}\ldots C_{n,t}, \label{coincidence-arrangement}
\eeqa
associated with functions $h_i$, $i=1,\ldots, n$, with mutually disjoint supports, converge, as $t\to +\infty$, on suitably chosen \emph{domains 
of Haag-Ruelle scattering states}\footnote{We consider only the limit $t\to+\infty$ and outgoing scattering states in this paper as the case $t\to-\infty$ is completely analogous.} (cf. Section~\ref{Haag-Ruelle-appendix}). The limit $\C_n^+$  can be interpreted as a coincidence arrangement of detectors which is sensitive to states containing a 
configuration of $n$ particles, with velocities in the supports of the functions $h_1,\ldots, h_n$.   

An important advance  was made  by Buchholz, who proved, for a sufficiently large class of observables $C$, the following bound:
\beqa
\sup_{t\in \real}\|C_t\one_{\Delta}(U)\|<\infty,
\eeqa
where $\one_{\Delta}(U)$ is the  projection on states whose energy-momentum belongs to a bounded Borel  set $\De$. 
(See \cite{Bu90} and Lemma~\ref{existence-lemma} below). This a priori estimate is a foundation
of the theory of particle weights \cite{BPS91, Po04.1, Po04.2,Dy10, DT11.1, DT11} and it 
implies, in particular, that the sequences $\C_{n,t}$ converge on \emph{all Haag-Ruelle scattering states of bounded energy}. However, the question of 
their convergence on the orthogonal complement of the subspace of scattering states, which is of crucial importance for the problem of  a complete 
particle interpretation of the theory (cf. Chapter 6 of \cite{Ha}), remained unanswered to date.   

In this paper we give a  solution of this problem in the case of $n=2$ for Araki-Haag  detectors~(\ref{coincidence-arrangement}) sensitive to 
massive neutral particles. More precisely, let $\ti p_1, \ti p_2$ be two energy-momentum vectors of massive particles.
We choose almost local observables $B_1$, $B_2$ whose energy-momentum transfers belong to small neighbourhoods of $-\ti p_1$, $-\ti p_2$,
respectively, and set $C_1:=B_1^*B_1$, $C_2:=B_2^*B_2$. Now let $\De$ be a small neighbourhood of $\ti p_1+\ti p_2$. Our main result is
the existence of 
\beqa
\C_{2}^+(\De):= \slim_{t\to +\infty} C_{1,t}C_{2,t} \one_{\Delta}(U) \label{detector-convergence}.
\eeqa
 Moreover, we show that the union of the ranges of all the operators $\C_{2}^+(\De)$, constructed as above, coincides with the subspace of two-particle Haag-Ruelle scattering states, whose total energy-momenta belong to~$\De$.   
This latter result, stated precisely in Thm.~\ref{Weak-asymptotic-completeness} below,  can be interpreted as a weak variant of 
two-particle asymptotic completeness. We point out that this generalized concept of complete particle interpretation does not imply  the
conventional one.

To illustrate this point, let us give a simple example of a theory which satisfies our general assumptions from Subsect. \ref{general-assumptions}
and is not asymptotically complete in the conventional sense: 
Let $\mco\mapsto \mfa(\mco)$ be the net of local algebras of  massive scalar free field theory acting on the Fock space $\mathcal F$ 
and let $U$ be the corresponding unitary representation of translations. 
Let $\mco\mapsto \mfa_{\mathrm{ev}}(\mco)$ be a subnet generated by even functions of the fields acting on the subspace 
$\mathcal F_{\mathrm{ev}}   \subset \mathcal F$ spanned by vectors with even particle numbers  
and let us set $U_{\mathrm{ev}}=U|_{\mathcal F_{\mathrm{ev} } }$. 
Then the net $\hat\mfa(\mco)=\mfa(\mco)\otimes \mfa_{\mathrm{ev}} (\mco)$, acting on $\mathcal F\otimes \mathcal F_{\mathrm{ev}}$ and
equipped with the  unitary representation of translations  $\widehat U=U\otimes U_{\mathrm{ev}}$, satisfies the assumptions from Subsect.~\ref{general-assumptions} but is not asymptotically complete in the conventional sense. In fact, the subspace $\Om\otimes \mathcal F_{\mathrm{ev}}$,
where $\Om$ is the Fock space vacuum, is orthogonal to all the Haag-Ruelle scattering states of the theory (except for the vacuum).  
In physical terms, this subspace describes `pairs of oppositely charged particles', whose mass hyperboloids do not appear 
in the vacuum sector. Due to the choice of the energy-momentum  transfers of $B_i$, the
asymptotic observables $\C_{2}^+(\De)$ annihilate such pairs of charged particles and, as stated in 
Thm.~\ref{Weak-asymptotic-completeness} below, only neutral particles remain in their ranges. 

We would like to stress that our result applies  to concrete interacting quantum field theories, as for example the $\la \phi^4_2$ model.
This theory is known to  possess a lower and upper mass gap at small coupling constants  $\la$, but its particle aspects  are rather
poorly understood. Asymptotic completeness is only known for total energies from the intervals
$[0,3m-\eps]$ and $[3m+\eps, 4m-\eps]$, where $m$ is the particle mass  and $\eps\to 0$ as $\la\to 0$  \cite{GJS73,SZ76,CD82}.
Since we can choose the region $\De$ in (\ref{detector-convergence}) outside of these intervals, our result provides
new information about the asymptotic dynamics of this theory.

Let us now describe briefly the main ingredients of the proof of existence of the limit  (\ref{detector-convergence}): 
Let $\C_{2,t}(\De)$ be the approximants on the r.h.s. of (\ref{detector-convergence}).
Exploiting locality and the disjointness of supports of $h_1,h_2$ one can write 
\beq\label{something}
\begin{array}{rl}
\C_{2,t}(\De)
=\int h_1\left(\frac{x_{1}}{t} \right)h_2\left(\frac{x_{2}}{t}\right) B^*_1(t,x_1)B_2^*(t,x_2)B_1(t,x_1)B_2(t,x_2)\one_{\Delta}(U)dx_{1}dx_{2} \\[2mm]
 +O(t^{-\infty}),
\end{array}
\eeq
 where $O(t^{-\infty})$ is a term tending to zero in norm faster than any inverse power of $t$. 
 In the next step we exploit our assumptions on the energy-momentum transfers of $B_1$, $B_2$,  which give for
any $\Psi\in \Ran\,\one_{\Delta}(U)$:
\beqa
B_1(t,x_1)B_2(t,x_2)\Psi=\Om(\Om|B_1(t,x_1)B_2(t,x_2)\Psi),
\eeqa
due to the presence of the  lower mass-gap. Thus we obtain
\beqa
\C_{2,t}(\De)\Psi=\int \H_t(x_1, x_2)F_t(x_1,x_2) B^*_1(t,x_1)B_2^*(t,x_2)\Om dx_{1}dx_{2}+O(t^{-\infty}),
 \label{sketch}
\eeqa
where 
\[
F_t(x_1,x_2):=(\Om|B_1(t,x_1)B_2(t,x_2)\Psi),\  \H_t(x_1,x_2):= h_1\left(\frac{x_{1}}{t}\right)
h_2\left(\frac{x_{2}}{t}\right).
\] 
We note that
by replacing $\H_t(x_1,x_2)F_t(x_1,x_2)$ in the first term on the r.h.s. of (\ref{sketch}) with $g_1(t,x_1)g_2(t,x_2)$, where $g_1,g_2$
are positive energy solutions of the Klein-Gordon equation, one would obtain a Haag-Ruelle scattering state (cf. Thm.~\ref{Haag-Ruelle}). 
While such replacement is not possible at finite times, it turns out that it can be performed asymptotically. In fact, 
Thm.~\ref{Main-result-half} below, reduces the problem of strong convergence of $t\mapsto \C_{2,t}(\De)$  to the existence of the following limit 
in the norm topology of $L^2(\real^{2d})$: 
\beqa
F_+:=\lim_{t\to\infty}\e^{\i t\tilde{\omega}(D_{\hx})} \H_tF_t, \label{sketch-limit}
\eeqa
where $\hx=(x_1,x_2)\in\real^{2d}$, $\tom(D_{\hx})= \omega(D_{x_{1}})+ \omega(D_{x_{2}})$ and $\om(k)=\sqrt{k^2+m^2}$ is the dispersion relation of the massive 
particles under study. 

A large part of our paper is devoted to the proof of existence of the limit (\ref{sketch-limit}). In the first step, 
taken in Lemma~\ref{almost-KG-equation}, we show that $F_t$ satisfies the following inhomogeneous evolution equation
\beqa
\pa_t F_t= -\i \tilde{\omega}(D_{\hx}) F_t+\R_t,\label{sketch-evolution}
\eeqa   
where, using locality, we show that  the  term $\R_t$ satisfies $\|\ti\H_t \R_t\|_{2}=O(t^{-\infty})$, for any 
$\ti{\H}_{t}(\hx):=\ti{\H}\left(\fr{\hx}{t}\right)$ with $\ti{\H}\in \coinf(\rr^{2d})$ vanishing near the diagonal $\{x_{1}= x_{2}\}$. Given (\ref{sketch-evolution}), we prove the existence of the limit 
(\ref{sketch-limit}) by extending the method of propagation estimates to inhomogeneous evolution equations. 

An important step is to obtain a {\em large velocity estimate}, for which the usual quantum mechanical proof does not apply, since in our case all propagation observables must vanish near the diagonal. Instead we use  a relativistic argument, based on the fact that hyperplanes $\{t= v\cdot x\}$ for $|v|>1$ are space-like (see Lemma \ref{3.3}). 
Another key ingredient is a {\em phase-space propagation estimate}, whose proof follows closely the usual quantum mechanical one. One new aspect, to which we will come back below, is the fact that the convex Graf function $R$ must now vanish near the diagonal.
  By combining the two 
propagation estimates in Prop~\ref{Klein-Gordon-approximation}, we obtain the existence of the limit (\ref{sketch-limit}) and 
therefore the convergence of Araki-Haag detectors~(\ref{detector-convergence}).   

It is a natural question if the convergence of $\C_{n,t}$ can also be shown   
for $n\neq 2$ by the methods described above. Perhaps surprisingly, this does not seem to be the case for  $n=1$, since it is 
difficult to filter out possible `pairs of charged particles' using only one detector (cf. the discussion above). 
However,  the situation looks much better for  
$n>2$. Here the initial steps of our analysis  can be carried out and difficulties arise only at the level of the phase-space propagation
estimate: The Graf function $R$ must  vanish not only near the diagonal $x_1=x_2$, but also near all the other collision planes 
$x_1=x_3$, $x_2=x_3$ etc. Since $R$ is supposed to be convex in some ball around the origin 
it must be zero in a neighbourhood of the convex hull of the collision planes restricted to this ball. Thus   a large 
and physically interesting part of the configuration space is out of reach of the phase-space propagation estimate for $n>2$.
It seems to us that new propagation estimates have to be developed to handle this problem.

We would like to point out that our analysis is closely related to quantum-mechanical scattering theory for dispersive systems (see e.g. \cite{Ge91,Zi97}). 
A simple example of a dispersive system is the following Hamiltonian
\beqa
H_{\mathrm{d}}=\sum_{i=1}^n\om(D_{x_i})+\sum_{i<j} V(x_i-x_j), \label{dispersive-Hamiltonian}
\eeqa
where $V\in \cS(\real^{d})$. 
We note that the corresponding Schr\"odinger equation  has the form
\beqa
\pa_t \Psi_t=-\i\sum_{i=1}^n\om(D_{x_i})\Psi_t-\i\sum_{i<j}V(x_i-x_j)\Psi_t, \label{Schroedinger-equation}
\eeqa
where $\Psi_t=\e^{-\i tH_{\mathrm{d}}}\Psi$, $\Psi\in L^2_{\mathrm{sym}}((\real^{d})^{\times n})$.
For $n=2$ equation~(\ref{Schroedinger-equation})  has a form of the evolution equation~(\ref{sketch-evolution}) with $F_t=\Psi_t$, and 
$\R_t=-\i V(x_1-x_2)\Psi_t$ which satisfies $\| G_t \R_t\|_2=O(t^{-\infty})$  as a consequence of the 
rapid decay of the potential. 
In the light of our discussion of equation~(\ref{sketch-evolution}), it is not a surprise that asymptotic completeness holds 
for dispersive systems for $n=2$, (which is actually a well known fact). However, the case $n>2$ is still open and requires new ideas. 

Our paper is organized as follows: In  Sect.~\ref{Framework} we recall the framework of local relativistic quantum field theory and 
state precisely  our results. In Sect. \ref{sec1} we introduce some notation  and terminology and collect the main properties of particle detectors.
In Sect. \ref{echi} we reduce the problem of convergence of the families of observables
(\ref{detector-convergence}) to the existence of the limit~(\ref{sketch-limit}) and   derive the inhomogeneous  evolution 
equation~(\ref{sketch-evolution}). In Sect. \ref{echa} we prove the convergence in (\ref{sketch-limit}) by showing large velocity and 
phase-space propagation estimates. 
In Sect. \ref{Haag-Ruelle-appendix} we recall some basic facts on the Haag-Ruelle scattering theory in the two-particle case.
The proof of Thm.~\ref{Weak-asymptotic-completeness}, which gives a weak form of two-particle asymptotic completeness, is presented in 
Sect. \ref{blit}. In Appendix~\ref{alo} we state some generalizations of standard abstract arguments to the inhomogeneous evolution equations. They are used in Section~\ref{echa}.
 
\vspace{0.5cm}

\noindent{\bf Acknowledgment:} W.D. would like to thank Detlev Buchholz for pointing out to him the problem of existence of particle detectors
 in relativistic QFT and  numerous interesting discussions. W.D. is also grateful for stimulating discussions with 
Jacob Schach M\o ller,  Alessandro~Pizzo and Wojciech De Roeck concerning the problem of existence  of asymptotic observables in non-relativistic QFT.
W.D. acknowledges financial support of the  German Research Foundation (DFG) within the stipend DY107/1--1 and 
hospitality of the Hausdorff  Research  Institute for  Mathematics, Bonn.

\section{Framework and results}\init\label{Framework}  

In this section we  recall the conventional framework of local quantum field theory and 
formulate precisely our main results.
\subsection{Nets of local observables}\label{general-assumptions}

As usual in the Haag-Kastler framework of local quantum field theory,
we consider   a  net  
\[
\mco\mapsto \mfa(\mco)\subset B(\hil)
\] 
of von Neumann algebras attached to open bounded regions of Minkowski space-time $\real^{1+d}$,
which satisfies the assumptions of isotony, locality, covariance w.r.t. translations, positivity of energy,
uniqueness of the vacuum and cyclicity of the vacuum. 

The assumption of \emph{isotony} says  that $ \mfa(\mco_1)\subset \mfa(\mco_2)$ if $\mco_1\subset\mco_2$.
It allows to define the $C^*$-inductive limit of the net, which  will be denoted by  $\mfa$.
\emph{Locality} means that $\mfa(\mco_1)\subset \mfa(\mco_2)'$ if $\mco_1$ and $\mco_2$ are space-like
separated. To formulate the remaining postulates, we  assume 
that  there exists  a strongly continuous unitary representation of translations 
\[
\real^{1+d}\ni (t,x)\mapsto U(t, x)=: \e^{\i (tH- x\cdot P)}\hbox{ on }\cH.
\]
We also introduce the group of automorphisms of $\mfa$ induced by $U$: 
\[
\alpha_{t,x}(B):= B(t,x):=U(t,x)BU^{*}(t,x), \ B\in \mfa, \ (t,x)\in \rr^{1+d}.
\]
The assumption of \emph{covariance} says that
\beqa
\alpha_{t,x}(\mfa(\mco))=\mfa(\mco+(t,x)), \ \forall \  \hbox{open bounded } \mco \hbox{ and }  (t,x)\in \rr^{1+d}.
\eeqa
We will need a restrictive formulation of positivity of energy, suitable for massive theories. 
We denote  by $H_m:=\{ (E,p)\in \real^{1+d}\setbar E=\sqrt{p^2+m^2}\}$ the mass hyperboloid of a particle of mass $m>0$
and set $G_{\mu}:=\{ (E, p)\in \real^{1+d}\setbar E\geq \sqrt{p^2+\mu^2}\}$.  We assume that:
\beq\label{spec-ass}
\begin{array}{rl}
i)&\Sp\, U=\{0\}\cup H_m\cup G_{\mu}\hbox{ for  some }m<\mu\leq 2m,\\[2mm]
ii)&\one_{\{0\}}(U)=|\Omega\rangle\langle \Omega|, \ \Omega\hbox{ cyclic for }\mfa. 
\end{array}
\eeq
The unit vector $\Omega$ will be called the {\em vacuum vector}, we denoted by $\Sp\, U\subset \rr^{1+d}$ 
the spectrum of $(H,P)$ and by $\one_{\Delta}(U)$ the spectral projection on a Borel set $\Delta\subset \rr^{d+1}$.  
Part $i)$ in (\ref{spec-ass})  encodes \emph{positivity of energy} and the presence of an upper and lower mass-gap.  
Part $ii)$ covers the \emph{uniqueness and cyclicity of the vacuum}.

\subsection{Relevant classes of observables} 

In this subsection we introduce some classes of observables, which enter into the formulation of our main results.
First, we recall the definition of \emph{almost local} operators. 
\begin{definition}
$B\in \mfa$  is  {\em almost local}  if  there exists a family $A_{r}\in \mfa(\mco_{r})$, where  $\mco_{r}:=\{x\in \rr^{1+d}\ : \ |x|\leq r\}$ 
is the double cone of radius $r$ centered at $0$,  s.t.  $\| B- A_{r}\| \in O(\langle r\rangle^{-\infty})$.
\end{definition}

To introduce another important class -- the \emph{energy-decreasing} operators -- we need some definitions: If $B\in \mfa$, we denote by  $\widehat{B}$ its Fourier transform:
\begin{equation}
\label{toto.e2}
\widehat{B}(E, p):= (2\pi)^{-(1+d)/2}\int\e^{-\i(Et-p\cdot x)}B(t,x)dtdx,
\end{equation}
defined as an operator-valued distribution.  We denote by $\supp(\widehat B)\subset \rr^{1+d}$ the support of $\widehat{B}$, called the {\em energy-momentum transfer of } $B$.  We recall the following well-known properties:
\begin{equation}
\label{toto.e3}
\begin{array}{rl}
i)& \widehat{\alpha_{t,x}(B)}(E, p)= \e^{\i (Et-p\cdot x)}\widehat{B}(E, p),\\[2mm]
ii)& \supp(\wh{B^{*}})= - \supp(\widehat B), \\[2mm]
iii)&B\one_{\Delta}(U)= \one_{\ov{\Delta+\supp(\widehat B)}}(U)B\one_{\Delta}(U), \ \forall \textrm{ Borel sets } \Delta\subset \rr^{1+d}.
\end{array}
\end{equation}
Now we are ready to define the energy-decreasing operators:
\begin{definition}
 $B\in \mfa$ is  {\em energy decreasing} if $\supp(\widehat B)\cap V_{+}=\emptyset$, where $V_{+}:= \{(E, p)\ : \ E\geq |p|\}$
is the closed forward light cone.
\end{definition}
\noindent In the rest of the paper we will work with the following set of observables:
\begin{definition}\label{def-de-L0}
 We denote by $\mcL_{0}\subset \mfa$ the subspace spanned by $B\in \mfa$ such that:
 \[
\begin{array}{rl}
i)&B\hbox{ is energy decreasing}, \ \supp(\widehat B)\hbox{ is compact},\\[2mm]
ii)&\rr^{1+d}\ni (t,x)\mapsto B(t,x)\in \mfa\hbox{ is }C^{\infty}\hbox{ in norm},\\[2mm]
iii)& \p^{\alpha}_{t,x}B(t,x) \hbox{ is almost local  for all }\alpha\in \nn^{1+d}. 
\end{array}
\]
\end{definition}
Note that if {\it i)} and {\it ii)} hold, then $\p_{t,x}^{\alpha}B(t,x)$ is energy decreasing for any $\alpha\in \nn^{1+d}$. Note also that if $A\in \mfa(\mco)$ and $f\in \cS(\rr^{1+d})$ with $\supp \widehat{f}$ compact and $\supp\widehat{f}\cap V_{+}=\emptyset$ then
\beq\label{arita}
B=(2\pi)^{-(1+d)/2} \int f(t,x)A(t,x)dtdx
\eeq
belongs to $\mcL_{0}$  by (\ref{toto.e3}) {\it i)}, since $\widehat{B}(E,p)= \widehat{f}(E, p)\widehat{A}(E,p)$. (See (\ref{toto.e1}) below for definition of $\widehat f$).

\subsection{Results}

For any $B_1,B_2\in \mcL_0$ and $h_1, h_2\in C_0^{\infty}(\real^{d})$ with disjoint supports we define the approximating families of {\em one-particle detectors}:
\beqa
C_{1, t}:=\int h_1\left(\fr{x_1}{t}\right)(B_1^*B_1)(t,x_1)dx_1,\quad  
C_{2,t}:=\int h_2\left(\fr{x_2}{t}\right)(B_2^*B_2)(t,x_2)dx_2\label{particle-detector-one}
\eeqa
which have appeared already  in (\ref{C-t-definition}) above. 
We note that in view of Lemma~\ref{existence-lemma}, stated below,
$\sup_{t\in\real}\|C_{i,t} \one_{\ti\Delta}(U)\|<\infty$, $i=1,2,$ for any bounded Borel set $\ti\De$. 

Now for any  open bounded subset $\De\subset G_{2m}$  we define  the {\em two-particle detectors}:
\beqa
\C_{2,t}(\De):=C_{1,t}C_{2,t} \one_{\Delta}(U).
\eeqa
Our main result is the strong convergence of $\C_{2,t}(\De)$ as $t\to\infty$ if the extension of $\De$ is smaller than the
mass-gap (i.e., $(\ov\De-\ov\De)\cap\Sp\, U=\{0\}$) and $(B_{1}, B_{2})$ is $\Delta-$admissible in the following sense: 
\begin{definition}\label{delta-admissible}
 Let $\Delta\subset \rr^{1+d}$ be an open bounded set 
 and $B_{1},B_{2}\in \mcL_{0}$. We say that $(B_{1}, B_{2})$ is $\Delta-${\em admissible} if 
 \beq\label{transfer-to-hyperboloid}
(-\supp(\widehat B_{i})) \cap \Sp\, U\subset H_m,\ i=1,2,
\eeq
\beq-(\supp(\widehat B_{1})+ \supp(\widehat B_{2}) )\subset \De, \label{transfer-from-vacuum}
\eeq
\beq(\ov{\De}+\supp(\widehat B_{1})+ \supp(\widehat B_{2}))\cap \Sp\, U\subset \{0\}. \label{transfer-to-vacuum}
\eeq
\end{definition}
\begin{remark}
In  Lemma~\ref{O-lemma}, it is shown that if  $\De\subset G_{2m}$ is an open bounded set s.t. $(\ov\Delta-\ov \Delta)\cap \Sp\, U\subset \{0\}$ and  $-\supp(\widehat B_{1})$, 
$-\supp(\widehat B_{2})$ are sufficiently small neighbourhoods 
of vectors $\tilde p_1,  \tilde p_2\in H_m$ s.t. $ \tilde p_1\neq  \tilde p_2$ and $  \tilde p_1+ \tilde p_2\in \De$ then $(B_{1}, B_{2})$ is $\Delta-$admissible.
\end{remark}
\bet\label{Main-theorem}  Let $\De\subset G_{2m}$ be an open bounded  set   such that $(\ov\De-\ov \De)\cap\Sp\, U=\{0\}$. Let $B_1, B_2\in  \mcL_0$ be $\De-$admissible and suppose that
$h_1,h_2\in C_0^{\infty}(\real^{d})$ have disjoint  supports.     Then there exists the limit
\beqa
\C_2^+(\De):=\slim_{t\to\infty}C_{1,t}C_{2,t} \one_{\Delta}(U), \label{main-approximants}
\eeqa
where $C_{i,t}$ are defined in  (\ref{particle-detector-one}) for  $B_i,h_i$, $i=1,2$. The range of $\C_2^+(\De)$
belongs to $\one_{\Delta}(U)\hil_2^+$, where $\hil_2^+$ is the subspace of two-particle Haag-Ruelle scattering states defined in Thm.~\ref{Haag-Ruelle}.
\eet
\proof  Follows immediately from Theorems~\ref{Main-result-half} and \ref{Klein-Gordon-approximation}. \qed

\medskip

Thm.~\ref{Main-theorem} is complemented by Thm.~\ref{Weak-asymptotic-completeness}, stated below,
which says that any two-particle scattering state can be prepared with the help of  Araki-Haag detectors. This weak variant of 
two-particle asymptotic completeness
ensures, in particular, that sufficiently many asymptotic observables~(\ref{main-approximants}) are non-zero.  
The proof is given in Sect. \ref{blit}. 
\bet\label{Weak-asymptotic-completeness}  Let $\De\subset G_{2m}$ be an open bounded set such that  $(\ov{\De}-\ov{\De})\cap\Sp\,U= \{0\}$. Let $J$ be the collection of quadruples $\al=(B_1,B_2,h_1,h_2)$ satisfying the conditions from Thm.~\ref{Main-theorem} and let $\C^+_{2,\al}(\De)$ be the 
limit (\ref{main-approximants}) corresponding to $\al$. Then 
\beqa
\one_{\Delta}(U)\hil_2^+=\Span\{  \Ran\, \C_{2,\al}^+(\De)\setbar     \al\in J  \}^{\cl}.  \label{weak-asymptotic-completeness}
\eeqa
\eet

\section{Preparations}\init\label{sec1}
In this section we introduce some notation  and collect some properties of particle detectors.
\subsection{Notation}\label{sec1.1}
\begin{enumerate}
\item[-] By $x,x_1,x_2$ we  denote elements of $\real^d$. We  set $\hx=(x_1,x_2)$ to denote elements of $\real^{2d}$.

\item[-] we write $K\Subset \rr^{1+d}$ if $K$ is a compact subset of $\rr^{1+d}$.

\item[-] we set $\langle x\rangle:= (1+ x^{2})^{\12}$ for $x\in \rr^{d}$ and $\omega(p)= (p^{2}+ m^{2})^{\12}$ for $p\in \rr^{d}$.

\item[-] the momentum operator $\i^{-1}\nabla_{x}$ will be denoted by $D_{x}$.

\item[-] we denote by $(t, \rx)$ or $(E, p)$ the elements of $\rr^{1+d}$.

\item[-] if $f: \rr^{1+d}\to \cc$ we will denote by $f_{t}: \rr^{d}\to \cc$ the function $f_{t}(\,\cdot\,):= f(t,\,\cdot\,)$.

\item[-] we denote by $\cS(\rr^{1+d})$ the Schwartz class in $\rr^{1+d}$. If $f\in \cS(\rr^{1+d})$ we define its (unitary) Fourier transform:
\begin{equation}
\label{toto.e1}
\widehat{f}( E, p):= (2\pi)^{-(1+d)/2}\int \e^{\i(Et-p\cdot x)}f(t,x)dt dx,
\end{equation}
so that
\beq
f(t,x)= (2\pi)^{-(1+d)/2}\int  \e^{-\i(Et-p\cdot x)}\widehat{f}(E, p)dEdp.
\eeq
Note the different sign in the exponent in
comparison with (\ref{toto.e2}).

If $f\in S(\rr^{d})$ we set:
\[
\widehat{f}(p)= (2\pi)^{-d/2}\int\e^{-\i p\cdot x}f(x)dx,
\]
and
\[
\widecheck{f}(x)= (2\pi)^{-d/2}\int \e^{\i p\cdot x }f(p)dp.
\]

\item[-] If $B$ is an observable, we write $B^{(*)}$ to denote either $B$ or $B^{*}$.  We will also set 
\[
B_{t}:= B(t, 0), \ B(x):= B(0,x)\hbox{ so that }B(t,x)= B_{t}(x).
\]

\end{enumerate}

\subsection{Auxiliary maps ${\mathbf a_B}$}\label{sec1.2}

For $B\in \mfa$, $f\in \cS(\rr^{d})$ we set:
\[
B(f):= \int B(x)f(x)dx,
\] 
so that $B^{*}(f)= B(\overline{f})^{*}$. Clearly, if $B_{1}, B_{2}\in \mfa$ are almost local, then
\begin{equation}
\label{quasi}
\| [B_{1}(x_{1}), B_{2}(x_{2})]\|\leq C_{N}\langle x_{1}- x_{2}\rangle^{-N}, \ \forall\ N\in \nn. 
\end{equation}
This immediately implies that
\begin{equation}
\label{quasi-1}
\|[B_{1}(f_{1}), B_{2}(f_{2})]\| \leq C_{N}\int |f_{1}(x_{1})|\langle x_{1}- x_{2}\rangle^{-N}|f_{2}(x_{2})| dx_{1}dx_{2}, \ f_{1}, f_{2}\in \cS(\rr^{d}).
\end{equation}
Now we introduce  auxiliary maps which will be often used  in our investigation: 
\bed
We denote by $a_{B}: \cH \to \cS'(\rr^{d}; \cH)$ the linear operator defined as:
\[
a_{B}\Psi(x):= B(x)\Psi, \ x\in \rr^{d}.
\]
\eed
\noindent Clearly $a_{B}: \cH \to  \cS'(\rr^{d}; \cH)$ is continuous and
\beq\label{toto.e01}
 B(f)= (\one_{\cH}\otimes\langle \overline{f}|)\circ a_{B}, \ f\in \cS(\rr^{d}),
\eeq
where $(\one_{\cH}\otimes\langle \overline{f}|): \cS'(\rr^{d}; \cH)\to \hil$ is defined on simple tensors by
\beqa
(\one_{\cH}\otimes\langle \overline{f}|)(\Psi\otimes T)=T(f)\Psi, \ \Psi\in \hil, T\in  \cS'(\rr^{d}).
\eeqa 
By duality $a_{B}^{*}: \cS(\rr^{d}; \cH)\to \cH$ is continuous and
\beq\label{toto.e02}
B^{*}(f)= a_{B}^{*}\circ (\one_{\cH}\otimes |f\rangle), \ f\in \cS(\rr^{d}).
\eeq
The group of space translations
\[
\tau_{y}\Psi(x):= \Psi(x-y), \ y\in\rr^{d},
\]
is a strongly continuous group on  $\cS'(\rr^{d}; \cH)$, and its generator is $D_{x}$ i.e.,
$\tau_{y}= \e^{- \i y\cdot D_{x}}$. 
It is easy to check the following identity:
\begin{equation}
\label{toto.e0}
a_{B}\circ\e^{-\i y\cdot P}= \e^{-\i y\cdot (D_{x}+ P)}\circ a_{B}, \ y\in \rr^{d}.
\end{equation}
We collect now some properties of $a_{B}$.
\begin{lemma}\label{newlemma}
 Let $B\in\mfa$. Then:
 \ben \item For any Borel set $\Delta\subset \rr^{1+d}$:
 \[
\begin{array}{rl}
a_{B}\one_{\Delta}(U)=&(\one_{\ov{\Delta+ \supp(\widehat B)}}(U) \otimes \one_{\cS'(\real^d)})\circ a_{B}\one_{\Delta}(U), \\[2mm]
a_{B}^{*}\circ(\one_{\Delta}(U)\otimes \one_{\cS(\real^d)})=&\one_{\ov{\Delta-\supp(\widehat B) }}(U)a_{B}^{*}\circ(\one_{\Delta}(U)\otimes \one_{\cS(\real^d)}).
\end{array}
\]
\item For any  $f\in \cS(\rr^{d})$ one has $f(D_{x})a_{B}= a_{B_{f}}$ for
\[
\begin{array}{rl}
B_{f}:=&(2\pi)^{-d/2}\int \widecheck{f}(-y)B(0, y)dy
 = (2\pi)^{-(d+1)/2}\int f(-p)\widehat{B}(E, p)dE dp,\\[2mm]
 \widehat{B_{f}}(E, p)=&f(-p)\widehat{B}(E, p).
 \end{array}
\]
\item If $\supp(\widehat B)$ is compact and $f\in C^{\infty}(\rr^{d})$ then the above properties also hold. 
\een
\end{lemma}
\proof (1) follows from (\ref{toto.e3}). (2) and (3) follow from the identity:
\[
\e^{- \i y\cdot D_{x}}a_{B}= a_{B(0, -y)}, \ y\in \rr^{d},
\]
which is a rephrasing of (\ref{toto.e0}). \qed

\medskip

If $B\in \mcL_{0}$, then $a_{B}$ has much stronger properties. In particular,   for $\Delta\Subset \rr^{1+d}$  
the operator $a_{B}\one_{\Delta}(U)$
maps $\cH$ into $L^{2}(\rr^{d}; \cH)\simeq \cH\otimes L^{2}(\rr^{d})$, see Lemma \ref{toto.1} below. 
This is a consequence  of the  following important property of  energy-decreasing operators, proven in~\cite{Bu90}.
\bel\label{existence-lemma} Let $B\in \mfa$ be energy-decreasing   with $\supp(\widehat B)\Subset \rr^{1+d}$ 
and $\De\subset\real^{1+d}$ be some bounded Borel set. Let $Y\subset\real^{1+d}$ be a subspace  and
let $dy$ be the Lebesgue measure on $Y$. Then there exists $c>0$ such that for any  $F\Subset Y$,  one has:
\beqa
\|\int_{F}  (B^*B)(y)\one_{\Delta}(U)dy\|\leq c\int_{F- F} \|[B^*,B(y)]\|dy. \label{harmonic}
\eeqa
\eel
Note that if $B$ is in addition almost local and $Y$ is spacelike, then the function $Y\ni y\mapsto \|[B^{*}, B(y)]\|$ vanishes faster than any inverse power of $|y|$ as $|y|\to \infty$, hence we can take $F=Y$ in (\ref{harmonic}).
We will usually apply this lemma with $Y=\{0\}\times \real^d$. 
In view of this lemma, it is convenient to introduce the subspace of vectors with compact energy-momentum spectrum:
\[
\cH_{\rm c}(U):= \{\Psi\in \cH \ : \ \Psi= \one_{\Delta}(U)\Psi, \ \Delta \Subset \rr^{1+d}\}.
\]
We note the following simple fact:
\begin{lemma}\label{toto.1} Assume that  $\Delta\Subset \rr^{1+d}$ and let $B\in \mcL_{0}$. Then
\[
a_{B}\one_{\Delta}(U): \cH\to \cH \otimes L^{2}(\rr^{d})\hbox{ is bounded.}
\] 
\end{lemma}
\begin{remark}
Considering  $a_{B}$ as a linear operator from $\cH$ to $ \cH\otimes L^{2}(\rr^{d})$ with domain $\cH_{\rm c}(U)$, we see that $\cH\otimes \cS(\rr^{d})\subset \Dom a_{B}^{*}$, hence $a_{B}$ is closable.
\end{remark}
\proof It suffices to note that
\[
\one_{\Delta}(U)a_{B}^{*}\circ a_{B}\one_{\Delta}(U)= \int_{\rr^{d}}\one_{\Delta}(U)(B^{*}B)(\rx) \one_{\Delta}(U) d \rx,
\]
and use  Lemma \ref{existence-lemma}.   \qed
\subsection{Particle detectors} In this subsection we make contact with the particle detectors $C_t$ introduced in (\ref{particle-detector-one}).
\begin{definition}\label{toto.2}
 Let $B\in \mcL_{0}$. For $h\in B(L^{2}(\rr^{d}))$ we set:
 \[
N_{B}(h):= a_{B}^{*}\circ(\one_{\cH}\otimes h)\circ a_{B}, \ \Dom N_{B}(h)= \cH_{\rm c}(U).
\]
\end{definition}
\noindent Denoting by $h(\rx,y)$ the distributional kernel of $h$ we have the  following  expression for $N_{B}(h)$,
\beq\label{toto.e1bis}
N_{B}(h)= \int B^{*}(x)h(x, y) B(y)d\rx dy,
\eeq
which makes sense as a quadratic form identity on $\cH_{\rm c}(U)$.
If $h$ is the operator of multiplication by the function $\rx\mapsto h(\rx)$, then $N_{B}(h)$ can be written as
\[
N_{B}(h)= \int (B^{*}B)(\rx)h(\rx)d \rx.
\]
Setting $h_{t}(x):= h\left(\xt\right)$, we see that $C_{t}$ 
defined in (\ref{particle-detector-one}) equals $N_{B_t}(h_{t})$, where $B_t=B(t,0)$. The following lemma is a direct consequence of  Lemmas \ref{newlemma} and  \ref{toto.1}.
\begin{lemma}\label{toto.3} We have:
\[
\begin{array}{rl}
(1)&\|N_{B}(h)\one_{\Delta}(U)\|_{B(\cH)}\leq c_{\Delta, B}\| h\|_{B(L^{2}(\rr^{d}))},\\[2mm]
(2)& \forall \ \Delta\Subset \rr^{1+d}, \ N_{B}(h)\one_{\Delta}(U)= \one_{\Delta_{1}}(U)N_{B}(h)\one_{\Delta}(U), \hbox{ for some }\Delta_{1}\Subset \rr^{1+d}.
\end{array}
\]
\end{lemma}

\medskip

\subsection{Auxiliary maps ${\mathbf a_{B_1,B_2}}$}\label{sec1.4}

 We start with the following definition which is meaningful
due to Lemma \ref{toto.1}:
\bed
If $B_{1}, B_{2}\in \mcL_{0}$, then we can define the linear operator:
\begin{equation}
\label{toto.e5}
a_{\BB}: \ \begin{array}{rl}
&\cH_{\rm c}(U)\to  \cH\otimes L^{2}(\rr^{2d},d \rx_{1}d\rx_{2}),\\[2mm]
&\Psi\mapsto a_{\BB}\Psi= (a_{B_{1}}\otimes \one_{L^{2}(\rr^{d})})\circ a_{B_{2}}\Psi.
\end{array}
\end{equation}
\eed
\noindent Formally we have 
\[
a_{\BB}\Psi(\rx_{1}, \rx_{2})= B_{1}(\rx_{1}) B_{2}(\rx_{2})\Psi.
\]
We note the following lemma, which is a direct consequence of Lemmas \ref{newlemma} and \ref{toto.1}.
\begin{lemma}\label{toto.-1}
 Assume $\Delta\subset \rr^{1+d}$ is compact and let $B_{1}, B_{2}\in \mcL_{0}$. Then:
\ben
\item $a_{\BB}\one_{\Delta}(U): \cH\to  \cH\otimes L^{2}(\rr^{2d},d \rx_{1}d\rx_{2})$ is bounded,

\item  for any $\Delta \Subset \rr^{1+d}$ one has:
\[
\begin{array}{rl}
&a_{\BB}\one_{\Delta}(U)=( \one_{\Delta+\supp(\widehat B_{1})+ \supp(\widehat B_{2}) }(U)\otimes \one_{L^{2}(\real^{2d}) })\circ a_{\BB}\one_{\Delta}(U), \\[2mm]
& a_{\BB}^{*}\circ ( \one_{\Delta}(U)\otimes \one_{L^{2}(\real^{2d})} )= 
\one_{\Delta-\supp(\widehat B_{1})-\supp(\widehat B_{2})}(U) a^{*}_{B_{1},B_{2}}\circ  ( \one_{\Delta}(U)\otimes \one_{L^{2}(\real^{2d}) }).
\end{array}
\]
\een
\end{lemma}

For later use we state in Lemma~\ref{stup} below a simple consequence of almost locality.  To simplify the formulation of this result, we
 introduce the following functions for $N >d$:
\beq\label{toto.e13bis}
g_{N}(k)= \int \e^{- \i k\cdot \rx} \langle \rx\rangle^{-N}d \rx.
\eeq
Clearly 
\[
\p_{k}^{\alpha}g_{N}(k)\in O(\langle k\rangle^{-p}), \ \forall \ p\in\nn, \ |\alpha|< N- |d|,
\]
and the operator on $L^{2}(\rr^{d})$ with kernel $\langle x- y\rangle^{-N}$ equals $g_{N}(D_{x})$.  
\begin{lemma}\label{stup}
 Let  $\Delta\Subset \rr^{1+d}$, $B_{i}\in \mcL_{0}$, $h_{i}\in C_{0}^{\infty}(\rr^{d})$, $i=1,2$.  We denote by $h_{i}\in B(L^{2}(\rr^{d}))$ the operator of multiplication by $h_{i}$. Then for any $N\in \nn$ one has:
 \beq\label{stupo}
\| \big(N_{B_{1}}(h_{1})\circ N_{B_{2}}(h_{2})- a_{\BBc}^{*}\circ (\one_{\cH}\otimes h_{1}\otimes h_{2})\circ a_{\BB}\big)\one_{\Delta}(U)\| 
\leq C_{N, \Delta, B_{1}, B_{2}} \| h_{1}g_{N}(D_{x})h_{2}\|_{ B(L^{2}(\real^{d})) }.
\eeq
\end{lemma}
\begin{remark}
In applications we will often estimate the operator norm on the r.h.s. of (\ref{stupo}) by the Hilbert-Schmidt norm 
$\|\,\cdot\,\|_{\mathrm{HS}}$.
\end{remark}

\proof Let $R$ be the operator in the l.h.s. of (\ref{stupo}). By Lemmas \ref{toto.3}, \ref{toto.-1}  $R= \one_{\Delta_{1}}(U)R\one_{\Delta_{2}}(U)$ for some $\Delta_{i}\Subset \rr^{1+d}$. For $u_{i}\in \cH$ we have
\[
\begin{array}{rl}
&|(u_{1}| Ru_{2})_{\cH}|\\[2mm]
=&|\int (\one_{\Delta_{1}}(U)u_{1}| B_{1}^{*}(x_{1}))[B_{1}(x_{1}), B_{2}^{*}(x_{2})]B_{2}(x_{2})h_{1}(x_{1})h_{2}(x_{2}) \one_{\Delta_{2}}(U)u_{2})_{\cH} dx _{1}dx_{2}|\\[2mm]
\leq &C\int  \| B_{1}(x_{1})\one_{\Delta_{1}}(U)u_{1}\|_{\cH} \| B_{2}(x_{2})\one_{\Delta_{2}}(U)u_{2}\|_{\cH} |h_{1}|(x_{1})|h_{2}|(x_{2})\langle x_{1}- x_{2}\rangle^{-N}dx_{1}dx_{2}.
\end{array}
\]
By Lemma \ref{toto.3} we know that $ v_{i}(x)= \|B_{i}(x)\one_{\Delta_{i}}(U)u_{i}\|_{\cH}\in L^{2}(\rr^{d})$ with $\| v_{i}\|_{L^{2}(\real^d)}\leq C_{i}\| u_{i}\|_{\cH}$.  Therefore 
\[
|(u_{1}| R u_{2})|_{\cH}\leq C \| |h_{1}| g_{N}(D_{x})|h_{2}| \|_{B(L^{2}(\real^d))}\| u_{1}\|_{\cH}\|u_{2}\|_{\cH}.
\]
Writing  $h_{i} = |h_{i}|{\rm sign}(h_{i})$ and using that the operator of multiplication by ${\rm sgn}(h_{i})$  is unitary, we obtain  the lemma. \qed

\section{An intermediate convergence argument}\label{echi}\init

For $B\in \mcL_{0}$ and $h\in \coinf(\rr^{d})$ we set:
\beq\label{def-de-Nb}
h_{t}(x):= h\left(\frac{x}{t}\right), \ N_{B}(h,t):= N_{B_{t}}(h_{t}).
\eeq
Recalling the notation $\hx=(x_1,x_2)$,  we also define $\tilde\omega(D_{\hx})= \omega(D_{x_{1}})+ \omega(D_{x_{2}})$, acting on $L^{2}(\rr^{2d})$.
The following theorem is an important step in the proofs of Thms. \ref{Main-theorem} and \ref{Weak-asymptotic-completeness}.  It essentially
 allows to  reduce their proofs to arguments adapted from non-relativistic scattering theory.
\bet\label{Main-result-half} 
Let $\Delta\subset \rr^{1+d}$ be a bounded open set,  $B_{1}, B_{2}\in \mcL_{0}$ with    $(B_{1}, B_{2})$ $\Delta-$admissible
and let $h_{1}, h_{2}\in\coinf(\rr^{d})$ have disjoint supports. Let
\beq
\H_{t}(x_1,x_2):= h_{1,t}(x_1) h_{2,t}(x_2)
\eeq
and  set for $\Psi\in \one_{\Delta}(U)\cH$:
\beq\label{defdeF-t}
F_{t}:= ( \langle \Omega|\otimes \one_{L^{2}(\real^{2d}) })\circ a_{\BB}\e^{-\i tH}\Psi\in L^{2}(\rr^{2d}), \,
\eeq
so that
\[
F_{t}(x_{1}, x_{2})= (\Omega| B_{1}(t, x_{1})B_{2}(t, x_{2})\Psi)_{\cH}, \ (x_{1}, x_{2})\in \rr^{2d}.
\]
Assume that:
\beq\label{tiri}
F_{+}:= \lim_{t\to \infty}\e^{ \i t\tilde{\omega}(D_{\hx})}\H_{t}F_{t}\hbox{ exists}.
\eeq
Then 
\beq\label{tara}
 \lim_{t\to \infty}N_{B_{1}}(h_{1}, t)N_{B_{2}}(h_{2}, t)\Psi
\eeq
exists and belongs to $\one_{\Delta}(U)\cH_{2}^{+}$.
\eet
\proof Applying  Lemma \ref{stup}  and noting that $\|h_{1,t}g_{N}(D_{x})h_{2,t}\|_{\rm HS}\in O(t^{d-N})$, we get:
\[
\begin{array}{rl}
N_{B_{1}}(h_{1, t})N_{B_{2}}(h_{2,t})\one_{\Delta}(U)= a_{\BBc}^{*}\circ (\one_{\cH}\otimes \H_{t})\circ a_{\BB}\one_{\Delta}(U)+ O(t^{-\infty}).
\end{array}
\]
By (\ref{transfer-to-vacuum}) and Lemma \ref{toto.-1} we have:
\[
 a_{\BB}\one_{\Delta}(U)=a ( \one_{\{0\}}(U)\otimes\one_{L^{2}(\real^{2d})})\circ a_{\BB}\one_{\Delta}(U)
 = ( |\Omega\rangle\langle \Omega|\otimes\one_{L^{2}(\real^{2d})  })\circ a_{\BB}\one_{\Delta}(U),
\]
using (\ref{spec-ass}). Therefore we have:
\beq\label{toto.e26}
\begin{array}{rl}
&\e^{\i tH}N_{B_{1}}(h_{1, t})N_{B_{2}}(h_{2,t})\e^{-\i tH}\Psi= \e^{\i tH}a_{\BBc}^{*} (\Omega\otimes \H_{t}F_{t}) + O(t^{-\infty})\\[2mm]
=& \e^{\i tH}a_{\BBc}^{*}(\Omega\otimes\e^{-\i t \tilde{\omega}(D_{\hx})}F_{+}) + o(t^{0}).
\end{array}
\eeq
Set
 \[
S_{t}: L^{2}(\rr^{2d})\ni F\mapsto  \ \e^{\i tH}a_{\BBc}^{*}(\Omega\otimes\e^{-\i t \tilde{\omega}(D_{\hx})}F) \in \cH.
\]
By Lemma \ref{toto.-1} the family $S_{t}$ is uniformly bounded in norm. Moreover if  $g_{1}, g_{2}$ are two positive energy KG solutions with disjoint velocity supports  (see Subsect. \ref{HR1} for terminology) and $f_{1}, f_{2}\in \cS(\real^d)$ are their initial data, then 
\[
S_{t}(f_{1}\otimes f_{2})= B^{*}_{1,t}(g_{1}) B^{*}_{2, t}(g_{2})\Omega,
\]
where the Haag-Ruelle creation operators $B^{*}_{i, t}(g_{i})$ are defined in Subsect. \ref{HR2}.  From Thm. \ref{Haag-Ruelle} we know that $\lim_{t\to \infty}S_{t}(f_{1}\otimes f_{2})$ exists. By linearity and density, using the uniform boundedness of $S_{t}$, we conclude that $\lim_{t\to\infty}S_{t}F$ exists for any $F\in L^{2}(\rr^{2d})$. By (\ref{toto.e26}) this implies the existence of the limit in (\ref{tara}).   The approximation argument above implies that this limit belongs to $\cH_{2}^{+}$.  The fact that it belongs to the range of $\one_{\Delta}(U)$ follows from the $\Delta-$admissibility of $(B_{1}, B_{2})$.\qed

\medskip

The proof of the existence of the limit (\ref{tiri}) will be given in the next section. As a preparation, we collect some properties of  the vectors $F_{t}\in L^{2}(\rr^{2d})$. The most important property is that $F_{t}$ solves a Schr\"{o}dinger equation with Hamiltonian $\tilde{\omega}(D_{\hx})$ and a {\em source  term} $\R_{t}$ whose $L^{2}$ norm outside of the diagonal decreases very fast when $t\to +\infty$.

\begin{lemma}\label{almost-KG-equation}
 Let $F_{t}$ be defined in (\ref{defdeF-t}). Then:
 \ben
 \item $F_{t}$ is uniformly bounded in $L^{2}(\rr^{2d})$,
 \item $t\mapsto F_{t}\in L^{2}(\rr^{2d})$ is $C^{1}$ with 
 \[
\p_{t}F_{t}= -\i \tilde{\omega}(D_{\hx})F_{t}+\R_{t}, 
\]
where $\|\ti\H\left(\frac{\hx}{t}\right)\R_{t}\|_{L^{2}(\real^{2d})}\in O(t^{-\infty})$ for any $\ti\H\in C_{0}^{\infty}(\rr^{2d})$ with $\supp \ti\H\cap \{x_{1}=x_{2}\}= \emptyset$.
 \een
\end{lemma}
\proof We have $F_{t}(x_{1}, x_{2})= (\Omega | B_{1}(t, x_{1})B_{2}(t, x_{2})\Psi)_{\cH}$ and from Lemma \ref{toto.-1} we know that $F_{t}$ is uniformly bounded in $L^{2}(\rr^{2d})$. Moreover, since $\Psi\in \cH_{\rm c}(U)$, we see that $t\mapsto F_{t}\in L^{2}(\rr^{2d})$ is $C^{1}$ with:
\[
\begin{array}{rl}
\p_{t}F_{t}= & (\Omega | \dot B_{1}(t, x_{1})B_{2}(t, x_{2})\Psi)_{\cH}+  (\Omega | B_{1}(t, x_{1})\dot B_{2}(t, x_{2})\Psi)_{\cH}\\[2mm]
=& (\Omega | \dot B_{1}(t, x_{1})B_{2}(t, x_{2})\Psi)_{\cH}+  (\Omega | \dot B_{2}(t, x_{2})B_{1}(t, x_{1})\Psi)_{\cH}\\[2mm]
&+  (\Omega | [B_{1}(t, x_{1}), \dot B_{2}(t, x_{2})]\Psi)_{\cH},
\end{array}
\]
where $\dot B_i:=\pa_s B_i(s,0)_{|s=0}$ are again almost local by the definition of $\mcL_0$. We have for any $\Phi\in \cH$:
\[
\begin{array}{rl}
(\Omega| B_{j}(t,x_{j})\Phi)_{\cH}= &(\Omega | \one_{\{0\}}(U)B_{j}(t,x_{j})\Phi)_{\cH}=(\Omega| B_{j}(t,x_{j})\one_{H_{m}}(U)\Phi)_{\cH}\\[2mm]
=&(\Omega| B_{j}(x_{j})\e^{-\i t \omega(P)}\Phi)_{\cH}= \e^{- \i t \omega(D_{x_{j}})}(\Omega |B_{j}(x_{j})\Phi),
\end{array}
\]
using (\ref{toto.e3}),  (\ref{transfer-to-hyperboloid}) and finally (\ref{toto.e0}). Differentiating this identity we obtain
\[
(\Omega| \dot B_{j}(t,x_{j})\Phi)_{\cH}= -\i\omega(D_{x_{j}}) (\Omega| B_{j}(t,x_{j})\Phi)_{\cH}.
\]
Therefore we get:
\[
\begin{array}{rl}
\p_{t}F_{t}=& -\i \omega(D_{x_{1}}) (\Omega | B_{1}(t, x_{1})B_{2}(t, x_{2})\Psi)_{\cH}-\i \omega(D_{x_{2}})(\Omega | B_{2}(t, x_{2})B_{1}(t, x_{1})\Psi)_{\cH}\\[2mm]
&+ (\Omega | [B_{1}(t, x_{1}), \dot B_{2}(t, x_{2})]\Psi)_{\cH}\\[2mm]
=&-\i \omega(D_{x_{1}})(\Omega | B_{1}(t, x_{1})B_{2}(t, x_{2})\Psi)_{\cH}-\i \omega(D_{x_{2}})(\Omega | B_{1}(t, x_{1})B_{2}(t, x_{2})\Psi)_{\cH}\\[2mm]
&-\i \omega(D_{x_{2}})(\Omega| [B_{2}(t, x_{2}), B_{1}(t, x_{1})]\Psi)_{\cH}+ (\Omega | [B_{1}(t, x_{1}), \dot B_{2}(t, x_{2})]\Psi)_{\cH}\\[2mm]
=&- \i \tilde{\omega}(D_{\hx})F_{t}+ \R_{t},
\end{array}
\]
for \[
\begin{array}{rl}
\R_{t}=& -\i \omega(D_{x_{2}})(\Omega| [B_{2}(t, x_{2}), B_{1}(t, x_{1})]\Psi)_{\cH}+ (\Omega | [B_{1}(t, x_{1}), \dot B_{2}(t, x_{2})]\Psi)_{\cH}\\[2mm]
=:& \R_{1,t}+ \R_{2,t}.
\end{array}
\]
Since  $\dot B_{2}$ is almost local, we have $\| [B_{1}(t, x_{1}), \dot B_{2}(t, x_{2})]\|\in O(\langle x_{1}- x_{2}\rangle ^{-N})$ uniformly in $t$ and $\| \ti\H_{t}\R_{2,t}\|_{L^{2}(\real^{2d}) }\in O(t^{-\infty})$ because of the support properties 
of $\ti\H_{t}(\hx):=\ti\H\left(\fr{\hx}{t}\right)$.

To estimate $\R_{1,t}$ we write it as $(\Omega| [\omega(D_{x_{2}})B_{2}(t, x_{2}), B_{1}(t, x_{1})]\Psi)_{\cH}$. By  Lemma \ref{newlemma} (2) we see that  $\omega(D_{x_{2}})B_{2}(t, x_{2})= C_{2}(t, x_{2})$, where 
\[
C_{2}= (2\pi)^{-d/2}\int f(x)B(0,x) dx, \ f\in \cS(\rr^{d}), \ \widehat{f}(-p)\equiv \omega(p)\hbox{ near }\supp(\widehat B_{2}).
\]
Therefore $C_{2}$ is almost local and $\| [C_{2}(t, x_{2}), B_{1}(t, x_{1})]\|\in O(\langle x_{1}- x_{2}\rangle^{-N})$. The same argument as above shows that 
$\|\ti\H_{t}\R_{1,t}\|_{L^{2}(\real^{2d})}\in O(t^{-\infty})$. \qed

\section{Non-relativistic scattering with source terms}\label{echa}\init
In this section we give the proof of the existence of the limit 
\[
F_{+}= \lim_{t\to +\infty}\e^{\i t \tilde{\omega}(D_{\hx})}\H_{t}F_{t},
\] appearing in Thm. \ref{Main-result-half}. The proof is obtained by adapting to our situation the standard arguments based on {\em propagation estimates}.
The main difference with the usual scattering theory is that $F_{t}$ solves a Schr\"{o}dinger equation with a {\em source term}.
This implies that one has to use  propagation observables supported in regions where the source term is small, in our case outside the diagonal in $\rr^{2d}$.   The necessary abstract arguments are collected in Appendix \ref{alo}.

\medskip

\subsection{Large velocity estimates}
In this subsection we prove large velocity estimates. Note that we do not prove them directly for $F_{t}$, but use instead a general argument based on Lemma \ref{existence-lemma}, locality and the fact that the hyperplanes $\{t= v\cdot x\}$ for $|v|>1$ are space-like.
\begin{lemma}\label{3.3}
 Let $B\in \mcL_{0}$, $\Delta\Subset \rr^{1+d}$ and $1<c<C$. Then,  
 \[
\int_{1}^{+\infty}(\e^{-\i tH}\Psi|\one_{\Delta}(U)N_{B}(\one_{\{z\in\real^d\ : \ c\leq |z|\leq C\}}\left(\frac{x}{t}\right))\one_{\Delta}(U)
\e^{-\i tH}\Psi)_{\cH}\frac{dt}{t}\leq c_{1}\|\Psi\|_{\cH}^{2}, \ \Psi\in \cH,
\]
where $x$ in the formula above denotes the corresponding multiplication operator on $L^2(\real^d)$.
\end{lemma}
\proof 
 Set $z=(z^{1}, z')\in\rr^{d}$ where $z^1\in\real$ is the first component of $z$. We can find constants $c_{i}>1$ and 
rotations $R_{i}\in SO(\rr^{d})$ such that
 \[
\begin{array}{rl}
\{z\ :\ c\leq |z|\leq C\}\subset \bigcup_{i=1}^{N}\{z\ : \ c_{i}\leq |(R_iz)^{1}|\leq C\}.
\end{array}
\]
 So it suffices to prove the lemma with $\one_{ \{  z \ : \   c\leq |z|\leq C\}}$ replaced with $\one_{\{ z \ : \    c\leq |(Rz)^{1}|\leq C\}}$ for $c>1$, $R\in SO(\rr^{d})$.  We parametrize the set $S= \{z\ : \ c_{i}\leq |(Rz)^{1}|\leq C\}$ by coordinates $(y^{1}, y')$ with $y^{1}= (Rx)^{1}$ so that it equals 
$S= \{(y^{1}, y')\ : \ c\leq |y^{1}|\leq C\}$. We have:
 \[
\begin{array}{rl}
I:=& \int_{1}^{\infty}\e^{\i tH}N_{B}(\one_{S}\left(\fr{x}{t} \right)   )\e^{-\i tH}\frac{dt}{t}=\int_{1}^{\infty}\frac{dt}{t}\int_{\rr^{d}}\one_{S}(\frac{y}{t})(B^{*}B)(t,y)dy\\[3mm]
=&\int_{1}^{\infty}dt\int_{c}^{C}dv\int_{\rr^{d-1}} (B^{*}B)(t, tv, y') dy'=\int_{c}^{C}dv \int_{\rr^{d}}(B^{*}B)(t, tv, y')dtdy'.
\end{array}
\]
 We now apply Lemma \ref{existence-lemma} to the subspace $Y_{v}=\{(t, tv, y')\ : \ t\in \rr, \ y'\in \rr^{d-1}\}$ for $c\leq v\leq C$ which yields:
 \beq\label{charia}
\|\one_{\Delta}(U)I \one_{\Delta}(U)\| \leq C'\int_{c}^{C}dv\int_{\rr^{d}}\| [B^{*}, B(t, tv, y')]\| dtdy'.
\eeq
 Since $B$ is almost local, there exist $B_{r}\in \mfa(\mco_{r})$ with $\|B- B_{r}\| \in O(\langle r\rangle^{-n})$. Therefore 
\[
\|[B^{*}, B(t, tv, y')]\|\leq C\langle r\rangle^{-n}+ \| [B^{*}_{r}, B_{r}(t, tv, y')]\|.
\]
Set $u\cdot u= x^{2}- t^{2}$ for $u= (t,x)\in \rr^{1+d}$. If $v_{1}, v_{2}\in\mco_{r}$ and $u_{1}= v_{1}+ (t, tv, y')$, $u_{2}= v_{2}$, then 
$u= u_{1}- u_{2}= (t, tv, y')+ w$, for $w\in \mco_{r}-\mco_{r}\subset \mco_{2r}$. It follows that 
\[
u\cdot u= t^{2}(|v|^{2}-1)+ |y'|^{2}+ O(r)(\langle t\rangle+\langle y'\rangle)+ O(r^{2}).
\]
Using that $c>1$,  we conclude  that there exists $0< \delta\ll 1$ such that if $\langle r\rangle = \delta(\langle t\rangle +\langle y'\rangle)$ 
then $\mco_{r}$ and $\mco_{r}+ (t, tv, y')$ are spacelike separated for any $(t,y')\in \rr^{d}$ s.t. $t^2+|y'|^2\geq 1$ and $c\leq v\leq C$. Therefore $\| [B^{*}, B(t, tv, y')]\|\in O(\langle t\rangle + \langle y'\rangle)^{-n}$, and the integral  in the r.h.s. of (\ref{charia}) is finite. \qed

To proceed we need the following definitions: For $0\leq r_{1}<r_{2}$ and $\epsilon\geq 0$ we set:
\[
C_{r_{1}, r_{2}}:=\{ \hx\in \rr^{2d}\ : \ r_{1}\leq |\hx| \leq r_{2}\}, \ C_{r}:= C_{0,r}, \ D_{\epsilon}:= \{\hx\in \rr^{2d}\ : \ | x_{1}- x_{2}|\leq \epsilon\}.
\]
Let us now prove the following corollary of Lemma \ref{3.3}:
\begin{proposition}\label{3.4}
 Let $\sqrt{2}<r<r'$, $\epsilon>0$ and let $F_{t}$ be defined in (\ref{defdeF-t}). Then
 \[
\int_{1}^{+\infty} \left\| \one_{C_{r,r'}\backslash D_{\epsilon}}\left(\fr{\hx}{t}\right)F_{t} \right\|^{2}_{L^2(\real^{2d}) }\frac{dt}{t}<\infty,
\]
where $\hx$ in the formula above denotes the corresponding multiplication  operator on $L^2(\real^{2d})$.
\end{proposition}
\proof 
Set  $\hx= (x_{1}, x_{2})\in \rr^{2d}$. 
By a covering argument, it suffices to prove the lemma with $\one_{C_{r,r'}\backslash D_{\epsilon}}(\hx)$ replaced with 
$h_{1}(x_{1})h_{2}(x_{2})$, where $h_{i}\in \coinf(\rr^{d})$ are supported near some points $y_i\in \real^{d}$ with  $(y_{1}, y_{2})\in C_{r,r'}\backslash D_{\epsilon}$ and ${\rm d}(\supp h_{1}, \supp h_{2})>0$.   Set $\H_{t}(\hx)= h_{1}(\frac{x_{1}}{t})h_{2}(\frac{x_{2}}{t})$. 
By (\ref{defdeF-t}) we have:
\[
(F_{t}| \H_{t}F_{t})_{L^{2}(\real^{2d}) }
=\int_{\rr^{2d}}(\e^{-\i tH}\Psi| B_{2}^{*}(x_{2})B_{1}^{*}(x_{1})B_{1}(x_{1})B_{2}(x_{2})\e^{-\i tH}\Psi)_{\cH}
h_{1}\left(\frac{x_{1}}{t}\right)h_{2}\left(\frac{x_{2}}{t}\right)dx_{1}dx_{2}.
\]
Since $|(y_{1}, y_{2})|> \sqrt{2}$, necessarily $|y_{i}|>1$ either for $i=1$ or $i=2$, and we can  assume that $\supp h_{i}\subset \{y\in \rr^{d}\ : \ |y|>1\}$.  If this holds for $i=2$ then
\[
\begin{array}{rl}
(F_{t}| \H_{t}F_{t})_{L^{2}(\real^{2d})}\leq &C \int(\e^{-\i tH}\Psi| B_{2}^{*}(x_{2})B_{2}(x_{2})\e^{- \i tH}\Psi)_{\hil}h_{2}(\frac{x_{2}}{t})dx_{2}\\[2mm]
\leq &C (\e^{-\i tH}\Psi|N_{B_{2}}(h_{2}(\frac{x }{t}))\e^{- \i tH}\Psi)_{\hil},
\end{array}
\]
where $x$ denotes the corresponding multiplication operator on $\real^{d}$.
Then we apply Lemma \ref{3.3}. If the above property holds for $i=1$ then using almost locality as in the proof of Lemma \ref{stup} we obtain that
\[
\begin{array}{rl}
&(F_{t}| \H_{t}F_{t})_{L^{2}(\real^{2d})}\\[2mm]
=&\int_{\rr^{2d}}(\e^{-\i tH}\Psi| B_{1}^{*}(x_{1})B_{2}^{*}(x_{2})B_{2}(x_{2})B_{1}(x_{1})\e^{-\i tH}\Psi)_{\cH}h_{1}(\frac{x_{1}}{t})h_{2}(\frac{x_{2}}{t})dx_{1}dx_{2}+ O(t^{-\infty})\\[2mm]
=&\int_{\rr^{d}}(B_{1}(x_{1})\e^{-\i t H}\Psi| N_{B_{2}}(h_{2}(\frac{x}{t}))B_{1}(x_{1})\e^{-\i tH}\Psi)_{\cH}h_{1}(\frac{x_{1}}{t})dx_{1}
+ O(t^{-\infty})\\[2mm]
\leq&C (\e^{-\i tH}\Psi|N_{B_{1}}(h_{1}(\frac{x}{t}))\e^{- \i tH}\Psi)_{\cH}+ O(t^{-\infty}),
\end{array}
\]
using that $h_{1}$, $h_{2}$ have disjoint supports. We complete the proof as before. \qed

\subsection{Phase-space propagation estimates}

We start with a geometrical consideration related to a well-known construction of Graf \cite{Gr90}.
\begin{lemma}\label{3.5}
 Let $K\Subset \rr^{2d}\backslash D_{0}$. Then there exist $\sqrt{2}<r<r'$, $c_{1}, c_{2}, \epsilon>0$ and a function $R\in \coinf(\rr^{2d})$ vanishing near $D_{0}$ such that
 \begin{equation}
\label{e5.1}
 \nabla^{2}R(\hx)\geq c_{1}\one_{K}(\hx)- c_{2}\one_{C_{r,r'}\backslash D_{\epsilon}}(\hx).
\end{equation}
\end{lemma}
\proof Set $\hx= (x_{1}, x_{2})\in \rr^{2d}$,  $u= \frac{1}{\sqrt{2}}(x_{1}+ x_{2})$, $v= \frac{1}{\sqrt{2}}(x_{1}- x_{2})$. We choose $\sqrt{2}<r<r'$ such that $K\subset C_{r}$ and set 
\[
 g(\hx)= (u^{2}+ \beta v^{2}-c)F(\hx),
\]
for $F\geq0$, $F\in \coinf(C_{r_{1}'})$, $F\equiv 1$ in $C_{r_{1}}$ where $r<r_{1}<r_{1}'<r'$.  The constants $c, \beta>0$ will be determined later. Note that $g$ is convex in $C_{r_{1}}$, hence
\[
R_{0}(\hx)= \sup\{g, 0\}(\hx)
\]
is convex in $C_{r_{1}}$ (but not smooth).  We first fix $c= r'^{2}$ so that $R_{0}(\hx)= 0$ for $\hx\in D_{\epsilon_{\beta}}$, for some $\epsilon_{\beta}>0$ tending to $0$ when $\beta\to +\infty$. We choose then  $\beta\gg 1$ such that $K\subset \{\hx\in \rr^{2d}\ : \ R_{0}(\hx)>0 \}$ and set $\epsilon= \epsilon_{\beta}$. By the continuity of $R_{0}$ we also obtain:
\begin{equation}
\label{e5.5}
K\subset\bigcap_{|\hx'|\leq \epsilon'}\{\hx\ : \ R_{0}(\hx-\hx')>0\},
\end{equation}
\begin{equation}
\label{e5.6}
D_{\epsilon/2}\subset\bigcap_{|\hx'|\leq \epsilon'}\{\hx\ : \ R_{0}(\hx-\hx')=0\},
\end{equation}
for some $\epsilon'\ll 1$. 

We now choose $\eta\geq 0$, $\eta\in \coinf(C_{\epsilon'})$ with $\int \eta(\hx)d\hx=1$ and set:
\[
R(\hx):= \int \eta(\hx')R_{0}(\hx-\hx')d\hx'= \eta\star R_{0}(\hx).
\]
Clearly $R\in \coinf(\rr^{2d})$ and $R$ is convex in $C_{r}$, hence
\begin{equation}
\label{e5.8}
\nabla^{2}R(\hx)\geq 0, \ \hx\in C_{r}.
\end{equation}
By relation~(\ref{e5.5}), $R= \eta\star g$ on $K$, hence
\begin{equation}
\label{e5.9}
\nabla^{2}R(\hx)\geq c_{1}\one, \ \hx\in K,
\end{equation}
for some $c_{1}>0$. In $C_{r,r'}$, $\nabla^{2}R$ is bounded, and outside of $C_{r'}$, $\nabla^{2}R(\hx)\geq 0$ since $R(\hx)\equiv 0$ there by construction. By (\ref{e5.8}), (\ref{e5.9}) we obtain (\ref{e5.1}). \qed

\medskip

\begin{proposition}
 \label{5.2}
Let $F_{t}$ be defined in (\ref{defdeF-t}) and $K\Subset \rr^{2d}\backslash D_{0}$. Then
\[
\int_{1}^{+\infty}\left\| \one_{K}\left(\frac{\hx}{t} \right)\left(\frac{\hx}{t}- \nabla \tilde{\omega}(D_{\hx})\right) F_{t}\right\|^{2}_{L^2(\real^{2d})} \frac{dt}{t}<\infty.
\]
\end{proposition}
\proof We will apply Lemma \ref{A1} to $\cH= L^{2}(\rr^{2d})$, $u(t)= F_{t}$, $H= \tilde{\omega}(D_{\hx})$ and 
\[
M(t)= R\left(\fr{\hx}{t}\right)- \12 \left(\nabla R\left(\fr{\hx}{t}\right)\cdot \left(\fr{\hx}{t}- \nabla \tilde{\omega}(D_{\hx})\right)+ \rm{h.c.}\right).
\]
Recall that $\D M(t)$ denotes the associated Heisenberg derivative.  By standard pseudo-differential calculus we obtain that:
\beq\label{e5.11}
\begin{array}{rl}
 \D M(t)= &\frac{1}{t}\left(\fr{\hx}{t} - \nabla \tilde{\omega}(D_{\hx})\right)\cdot 
 \nabla^{2}R\left(\fr{\hx}{t}\right)  \cdot \left( \fr{\hx}{t}- \nabla \tilde{\omega}(D_{\hx})\right) + O(t^{-2})\\[2mm]
 \geq &\frac{c_{1}}{t} \left( \fr{\hx}{t}  - \nabla \tilde{\omega}(D_{\hx})\right)\one_{K}\left(\fr{\hx}{t}\right) \cdot \left( \fr{\hx}{t}- \nabla \tilde{\omega}(D_{\hx})\right)- 
\frac{C}{t}\one_{C_{r,r'}}\left(\fr{\hx}{t} \right)+ O(t^{-2}),
\end{array}
\eeq
where $O(t^{-2})$ denotes a term with norm $O(t^{-2})$ and we have used Lemma \ref{3.5} in the second line. 
Since $R$ is supported away from the diagonal, we obtain by Lemma \ref{almost-KG-equation} and pseudo-differential calculus that 
$\|M(t) \R_{t}\|\in L^{1}(\real^+,dt)$, where we recall that $\p_{t}F_{t}=: -\i \tilde{\omega}(D_{\hx})F_{t}+ \R_{t}$. Lemma~\ref{almost-KG-equation} also gives that $\sup_{t}\| F_{t}\|<\infty$. The negative term in the r.h.s. of (\ref{e5.11}) is controlled by Proposition~\ref{3.4}. Applying Lemma~\ref{A1} we obtain the desired result.\qed 

\subsection{Existence of the intermediate limit}

\begin{theoreme}\label{Klein-Gordon-approximation}
 Let $F_{t}$, $\H_{t}$ be defined in (\ref{defdeF-t}). Then the limit
 \[
F_{+}= \lim_{t\to +\infty}\e^{\i t \tilde{\omega}(D_{\hx})}\H_{t}F_{t}\hbox{ exists}.
\]
\end{theoreme}
\proof  All the norms and scalar products in this proof are in the sense of  $L^2(\real^{2d})$.
We proceed as in the proof of \cite[Prop. 4.4.5]{DG97}. Set first $\H(\hx)= h_{1}(x_{1})h_{2}(x_{2})$ and
\[
M(t)= \H\left( \fr{\hx}{t}  \right)- \left(\fr{\hx}{t}- \nabla \tilde{\omega}(D_{\hx}) \right)\cdot \nabla \H\left( \fr{\hx}{t}  \right).
\]
By pseudo-differential calculus, we obtain that
\beq\label{e6.1}
\begin{array}{l}
\D M(t)= \frac{1}{t}\left( \fr{\hx}{t}- \nabla \tilde{\omega}(D_{\hx})\right)\cdot 
 \nabla^{2}\H\left( \fr{\hx}{t}  \right)\cdot \left( \fr{\hx}{t}   -\nabla \tilde{\omega}(D_{\hx})\right) + O(t^{-2}),\\[2mm]
  \| M(t)\R_{t}\|, \| M^{*}(t)\R_{t}\| \in L^{1}(\rr^+, dt),
\end{array}
\eeq
where in the second line we use that $\H$ is supported away from the diagonal.  Note that the following analog  of Prop. \ref{5.2} is well-known and easy to prove by mimicking the arguments in \cite[Prop. 4.4.3]{DG97}: 
\begin{equation}
\label{e6.2}
\int_{1}^{+\infty}  \left\| \one_{K}\left( \fr{\hx}{t} \right)\left( \fr{\hx}{t}- \nabla \tilde{\omega}(D_{\hx})  \right)
\e^{- \i t \tilde{\omega}(D_{\hx})}u\right\|^{2}\frac{dt}{t}\leq C \| u\|^{2}, \ u\in L^{2}(\rr^{2d}),
\end{equation}
for any $K\Subset \rr^{2d}\backslash \{0\}$. Combining this estimate with the one in Prop. \ref{5.2}, we obtain 
by Lemma~\ref{A2}  that
\[
\lim_{t\to +\infty}\e^{\i t \tom(D_{\hx})}M(t)F_{t}\hbox{ exists}.
\]
Therefore the proposition follows if we show that
\[
\lim_{t\to \infty}\left( \fr{\hx}{t} - \nabla \tilde{\omega}(D_{\hx})\right)\cdot \nabla \H\left( \fr{\hx}{t} \right)  F_{t}=0,
\]
or equivalently 
\beq\label{e6.3}
\lim_{t\to +\infty} (F_{t}| \left( \fr{\hx}{t}-\nabla \tilde{\omega}(D_{\hx}) \right)\ti G\left(\fr{\hx}{t}\right)\left(\fr{\hx}{t}-\nabla \tilde{\omega}(D_{\hx}) \right)F_{t})_{L^2(\real^{2d}) }=0,
\eeq
for $\ti G=\ti H\one$,  $ \ti H  \in \coinf(\rr^{2d}\backslash D_{0})$ and $\ti H\geq 0$.  It suffices to prove that the limit in (\ref{e6.3}) exists, 
since it will then be equal to $0$ by Prop. \ref{5.2}. To this end, we apply Lemma~\ref{A3} with
\[
M(t)= \left(\fr{\hx}{t}  - \nabla \tilde{\omega}(D_{\hx})  \right)\ti G\left( \fr{\hx}{t}  \right)\left(\fr{\hx}{t} - \nabla\tilde{\omega}(D_{\hx})\right).
\]
Again $\|M(t)\R_{t}\|, \| M^{*}(t)\R_{t}\|\in L^{1}(\rr^{+},dt)$ and by pseudo-differential calculus we have:
\[
\begin{array}{rl}
\D M(t)=& -\fr{2}{t}\left(\fr{\hx}{t}- \nabla \tilde{\omega}(D_{\hx}) \right)\ti G\left(  \fr{\hx}{t} \right) \left( \fr{\hx}{t}  - \nabla \tilde{\omega}(D_{\hx}) \right)\\[2mm]
&-\fr{1}{t}\left( \fr{\hx}{t} - \nabla \tilde{\omega}(D_{\hx}) \right)\nabla \ti G\left(\fr{\hx}{t}  \right)\cdot \left( \fr{\hx}{t}-\nabla \tilde{\omega}(D_{\hx})\right)\left(\fr{\hx}{t}  -\nabla \tilde{\omega}(D_{\hx})\right)+ O(t^{-2})\\[2mm]
=& \fr{1}{t}\left( \fr{\hx}{t}- \nabla \tilde{\omega}(D_{\hx}) \right) \one_{K}\left(\frac{\hx}{t} \right)  A(t) \one_{K}\left(\frac{\hx}{t} \right)\left( \fr{\hx}{t}- \nabla \tilde{\omega}(D_{\hx})\right)+ O(t^{-2}),
\end{array}
\]
for a compact set $K\subset \real^{2d}\backslash D_0$ and $A(t)\in O(1)$. Now the existence of the limit follows  from Prop. \ref{5.2} and Lemma \ref{A3}. \qed

\section{Haag-Ruelle scattering theory}\init\label{Haag-Ruelle-appendix}
In this section we recall some basic facts concerning the Haag-Ruelle scattering theory.

\subsection{Positive energy solutions of the Klein-Gordon equation}\label{HR1}
\begin{definition} \label{KG-definition}
 Let  $f\in\cS(\rr^{d})$,  such that $\widehat{f}$ has compact support. The function 
\[
g(t,x)= g_{t}(x)\hbox{ for }g_{t}= \e^{-\i t \omega(D_{x})}f,
\]
which solves $( \p_{t}^{2}- \Delta_{x}) g + m^{2}g=0$, 
will be called a  {\em positive energy KG solution}.
\end{definition}
\bep There hold the following facts:  \label{toto.20}
 \ben 
 \item Let $h\in \coinf(\rr^{d})$. Then  
 \[
\slim_{t\to \pm \infty}\e^{\i t \omega(D_{x})}h\left(\frac{x}{t}\right)\e^{- \i t \omega(D_{x})}= h(\nabla\omega(D_{x})).
\]
\item Let $\chi_{1}, \chi_{2}\in C^{\infty}(\rr^{d})$ be bounded with all derivatives and having disjoint supports. Let $f\in\cS(\rr^{d})$
be s.t. $\widehat f$ has compact support. Then
\[
\|\chi_{1}\left(\frac{x}{t}\right)\e^{- \i t \omega(D_{x})}\chi_{2}(\nabla \omega(D_{x}))f\|_{L^2(\real^d)} \in O(t^{-\infty}).
\]
\een
\eep
\proof (1) is obvious. For (2) see \cite{RS3}. \qed\\
The following notion of  {\em velocity support} will be useful later on.
\begin{definition}\label{def-de-vel}
 Let $\Delta\Subset H_{m}$. We set
 \[
{\rm Vel}(\Delta):= \{\nabla \omega(p)\ : \ p\in \rr^{d}, \ (\omega(p), p)\in \Delta\}.
\]
\end{definition}
\noindent Clearly if $\Delta_{1}$ and $\Delta_{2}$ are disjoint, then  so are ${\rm Vel}(\Delta_{1})$ and ${\rm Vel}(\Delta_{2})$.
If $g$ is a positive energy KG solution with initial data $f$, then $\supp\widehat{g}\subset H_{m}$ and 
${\rm Vel}(\supp \widehat{g})= \{\nabla \omega(p) \setbar    p\in \supp \widehat{f}\}$ can be called the {\em velocity support} of $g$, as illustrated 
by Prop.~\ref{toto.20} (2). 

\subsection{Haag-Ruelle scattering theory}\label{HR2}
Let $B\in  \mcL_{0}$ satisfy (\ref{transfer-to-hyperboloid}), i.e.,
\[
-\supp(\widehat B)\cap \Sp\, U\subset H_{m}.
\] Let now $g$ be a positive energy KG solution. The {\em Haag-Ruelle creation operator} 
is given by $\{B^{*}_{t}(g_t)\}_{t\in \rr}$, that is,
\[
 B_{t}^{*}(g_{t})= \int g(t,x)B^{*}(t,x)dx.
\]
Note that since $\e^{- \i t \omega(D_{x})}$ preserves $\cS(\rr^{d})$ the integral is well defined.

\begin{lemma}\label{toto.21}
 The following properties hold:
 \ben
 \item $B_{t}^{*}(g_t)\Omega= B^{*}(f)\Omega= (2\pi)^{d/2}\widehat{f}(P)B \Omega$, if $g_{t}= \e^{-\i t \omega(D_{x})}f$.
 \item Let $\Delta\Subset \rr^{1+d}$, $f\in L^{2}(\rr^{d})$. Then $\| B^{(*)}(f)\one_{\Delta}(U)\|\leq c_{\Delta, B}\| f\|_{L^{2}(\rr^{d})}$.
 \item $\p_{t}B_{t}^{*}(g_t)= \dot{B}^{*}_{t}(g_t)+ B_{t}^{*}(\dot{g}_t)$, where $\dot{B}= \p_{s}B(s, 0)_{\mid s=0}\in \mcL_{0}$ and $\dot{g}= \p_{t}g$ is a positive energy KG solution with the same velocity support as $g$.
 \een
\end{lemma}
\proof 
We use the notation from Subsect. \ref{sec1.2}.   We have
\[
\begin{array}{rl}
B_{t}^{*}(g_t)\Omega=&(\langle \overline{g_{t}}|\otimes \one)\circ a_{B^{*}_{t}}\Omega= (\langle \overline{g_{t}}|\otimes \one)\circ(\one\otimes \e^{\i t H})\circ  a_{B^{*}}\Omega.
\end{array}
\]
By (\ref{transfer-to-hyperboloid}) and (\ref{toto.e3}) {\it iii)} we have $a_{B^{*}}\Omega= (\one\otimes\one_{H_{m}}(U))\circ a_{B^{*}}\Omega$, hence
\[
(\one\otimes \e^{\i tH})\circ a_{B^{*}}\Omega= (\one\otimes \e^{\i t\omega(P)})\circ a_{B^{*}}\Omega.
\]
From (\ref{toto.e0}) we obtain that:
\begin{equation}
\label{toto.e20}
(\one\otimes\e^{- \i y\cdot P})\circ a_{B^{*}}\Omega= (\e^{\i y\cdot D_{x}}\otimes \one)\circ a_{B^{*}}\Omega, \ y\in \rr^{d},
\end{equation}
which implies that 
\[
(\one\otimes \e^{\i t\omega(P)})\circ a_{B^{*}}\Omega= (\e^{\i t \omega(D_{x})}\otimes \one)\circ a_{B^{*}}\Omega,
\]
using that $\omega(p)= \omega(-p)$. Hence 
\[
\begin{array}{rl}
B_{t}^{*}(g_t)\Omega
=&(\langle \overline{g_{t}}|\otimes \one)\circ(\e^{\i t \omega(D_{x})}\otimes \one)\circ  a_{B^{*}}\Omega=(\langle \e^{-\i t \omega(D_{x})}\overline{g_{t}}|\otimes \one)\circ  a_{B^{*}}\Omega\\[2mm]
=&(\langle \overline{f}|\otimes \one)\circ a_{B^{*}}\Omega=B^{*}(f)\Omega.
\end{array}
\]
The fact that $B^{*}(f)\Omega= (2\pi)^{d/2}\widehat{f}(P)B^{*}\Omega$ is immediate. 
Statement (2) follows from Lemma \ref{toto.1}, using (\ref{toto.e01}) for $B$ and (\ref{toto.e02}) for $B^{*}$. In the  case of $B^*$ we 
also use Lemma~\ref{newlemma} (1) and the fact that
$\supp(\widehat B)$ is compact. (3) is a trivial computation. \qed

\medskip

The following  result  is a special case of the  Haag-Ruelle theorem \cite{Ha58, Ru62}. 
For the reader's convenience we give an elementary proof which combines ideas from \cite{BF82, Ar99, Dy05} and exploits the bound (2) in Lemma \ref{toto.21}.

\bet\label{Haag-Ruelle} Let $B_1, B_{2}\in \mcL_{0}$   satisfy (\ref{transfer-to-hyperboloid}). Let $g_1,g_2$ be
two positive energy KG solutions with disjoint velocity supports. Then: 
\ben
\item  There exists the two-particle scattering state given by
\beqa
\Psi^+=\lim_{t\to\infty} B^{*}_{1,t}(g_{1,t})B^{*}_{2,t}(g_{2,t})\Om. \label{scattering-state}
\eeqa
\item The state $\Psi^+$ depends only on the single-particle vectors $\Psi_i=B^{*}_{i,t}(g_{i,t})\Om$,  and therefore we can write $\Psi^+=\Psi_1\timeso\Psi_2$.  Given two such vectors $\Psi^+$ and $\tilde \Psi^+$ one has:
\beqa
(\tilde \Psi^+|\Psi^+)&=&( \tilde \Psi_1|  \Psi_1)  ( \tilde\Psi_2| \Psi_2)+( \tilde\Psi_1|  \Psi_2)  ( \tilde\Psi_2| \Psi_1),  
     \label{scalar-product}\\
U(t,x)(\Psi_1\timeso\Psi_2)&=&(U(t,x)\Psi_1)\timeso(U(t,x)\Psi_2), \ (t,x)\in \rr^{1+d}. \label{energy-factorization-relation}
\eeqa
\een
\eet
Before giving the proof of the theorem, let us explain how to obtain two-particle scattering states from arbitrary one-particle states, thereby defining the (outgoing) {\em two-particle wave operator}. Let  
\[
\cH_{m}:= \one_{H_{m}}(U)\cH,
\]
be the space of {\em one-particle states}.  
For $\Psi_{1}, \Psi_{2}\in \cH$ we set
\[
 \Psi_{1}{\otimes}_{\rm s}\Psi_{2}:= \frac{1}{\sqrt{2}}(\Psi_{1}\otimes \Psi_{2}+ \Psi_{2}\otimes\Psi_{1})\in \cH\otimes_{\rm s}\cH.
\]
\begin{proposition}
 \label{def-de-wave}
 There exists a unique isometry
 \[
W_{2}^{+}: \ \cH_{m}\otimes_{\rm s} \cH_{m}\to \cH
\] 
with the following properties:
\ben
\item If $\Psi_{1}$, $\Psi_{2}$ are as in Thm. \ref{Haag-Ruelle}, then $W_{2}^{+}(\Psi_{1}{\otimes}_{\rm s}\Psi_{2})= \Psi_{1}\timeso\Psi_{2}$, \vspace{1mm}
\item 
$U(t,x) \circ W_{2}^{+}= W_{2}^{+} \circ (U_m(t,x)\otimes U_m(t,x)), \ (t,x)\in \rr^{1+d}$, 
where we  denote by $U_m(t,x)$ the restriction of $U(t,x)$ to $\cH_{m}$.
\een
\end{proposition}
\begin{definition}
\ben
\item The map $W_{2}^{+}: \  \cH_{m}\otimes_{\rm s} \cH_{m}\to \cH$ is called the {\em (outgoing) two-particle wave operator}. 
\item   The range of $W_{2}^{+}$ is denoted by $\cH_{2}^{+}$.
\een
\end{definition}
\noindent {\em Proof of Prop. \ref{def-de-wave}.} Let us denote by ${\mathcal F}\subset \cH_{m}\otimes_{\rm s}\cH_{m}$ the subspace spanned by the vectors $\Psi_{1}{\otimes}_{\rm s}\Psi_{2}$ for $\Psi_{1}$, $\Psi_{2}$ as in Thm. \ref{Haag-Ruelle}.  By (\ref{scalar-product})
 there exists a unique isometry $W_{2}^{+}: {\mathcal F}\to \cH$ such that 
 \[
W_{2}^{+}(\Psi_{1}{\otimes}_{\rm s}\Psi_{2})= \Psi_{1}\timeso \Psi_{2},
\]
for all $\Psi_{1}$, $\Psi_{2}$ as in the theorem. Moreover  by (\ref{energy-factorization-relation}) 
$U(t,x)\circ W_{2}^{+}= W_{2}^{+} \circ (U_m(t,x)\otimes U_m(t,x))$.  To complete the proof of the proposition it suffices to prove that the closure of ${\mathcal F}$ is $\cH_{m}\otimes_{\rm s}\cH_{m}$. 

Denote by $(H_{1}, P_{1})$, resp. $(H_{2}, P_{2})$ the generators of the groups $U_m(t,x)\otimes\one$, resp. $\one\otimes U_m(t,x)$ acting on $\cH_{m}\otimes \cH_{m}$, and set $(\tilde{H}, \tilde{P}):= ((H_{1}, P_{1}), (H_{2}, P_{2}))$, whose joint spectral measure is supported by $H_{m}\times H_{m}$.

 By Lemma \ref{toto.21} (1) and  the cyclicity of the vacuum, the set of vectors $B^{*}_{t}(g_t)\Om$, for $B\in\mcL_{0}$ satisfying (\ref{transfer-to-hyperboloid}) and $g$ a positive energy KG solution, is dense in $\cH_{m}$.
 Moreover for $\Delta\Subset H_{m}$, the set of such vectors  with $g$ having the velocity support included in ${\rm Vel}(\Delta)$ is dense in $\one_{\Delta}(U)\cH_{m}$. It follows  from these density properties that the closure of ${\mathcal F}$ in $\cH_{m}\otimes_{\rm s}\cH_{m}$ equals  
\[
 {\mathcal F}^{\rm cl}= \Theta_{\rm s} \circ \one_{(H_{m}\times H_{m})\backslash D}(\tilde{H}, \tilde{P})(\cH_{m}\otimes \cH_{m}),
 \]
 where $\Theta_{\rm s}: \cH_{m}\otimes \cH_{m}\to \cH_{m}\otimes_{\rm s}\cH_{m}$ is the orthogonal projection, and $D\subset H_{m}\times H_{m}$ is the diagonal.
 From \cite[Prop. 2.2]{BF82} we know that the spectral measure of  the restriction of $(H, P)$ to $\cH_{m}$ is absolutely continuous w.r.t. the Lorentz invariant measure on $H_{m}$. This implies that  $\one_{D}(\tilde{H}, \tilde{P})=0$, which completes the proof of the proposition. \qed

\medskip

\noindent {\em Proof of Thm. \ref{Haag-Ruelle}.}  Let us first prove (1). Let $B_{1}, B_{2}, g_1,g_2$ satisfy the hypotheses of the theorem. 
We claim that
\begin{equation}
\label{toto.e22}
 [B^{(*)}_{1,t}(g_{1,t}), B^{(*)}_{2,t}(g_{2,t})] \in O(t^{-\infty}).
\end{equation}
In fact by Prop. \ref{toto.20} (2)  
we can find cutoff functions $\chi_{1}, \chi_{2}\in \coinf(\rr^{d})$ with disjoint supports such that
\[
g_{i, t}= \chi_{i}\left(\frac{x}{t}\right)g_{i,t} + O(t^{-\infty})\hbox{ in }L^{2}(\rr^{d}).
\]
Setting $\chi_{i,t}(x)= \chi_{i}(\frac{x}{t})$, this implies by Lemma \ref{toto.21} (2) that:
\[
[B^{(*)}_{1,t}(g_{1,t}), B^{(*)}_{2,t}(g_{2,t})]= [B^{(*)}_{1,t}(\chi_{1,t}g_{1,t}), B^{(*)}_{2,t}(\chi_{2,t}g_{2,t})]+ O(t^{-\infty}).
\]
By the almost locality of $B^{(*)}_{1}, B^{(*)}_{2}$ we obtain from (\ref{quasi-1}) and the Cauchy-Schwarz inequality
that the commutator in the r.h.s. is bounded by 
\[
 C_{N}\|\chi_{1,t}g_{N}(D_{x})\chi_{2,t}\|_{\mathrm{HS} }\| g_{1, t}\|_{L^{2}(\real^d)  }\| g_{2,t}\|_{L^{2}(\real^d)}\in O(t^{-\infty}),
 \]
 which proves (\ref{toto.e22}). (Cf. the proof of Lemma~\ref{stup}).  Now  we get that
\[
\p_{t}(B_{1,t}^{*}(g_{1,t}))B_{2,t}(g_{2,t}))\Omega= [\p_{t}B_{1,t}^{*}(g_{1,t}), B_{2,t}^{*}(g_{2,t})]\Omega\in O(t^{-\infty}),
\]
where we made use of  Lemma \ref{toto.21} (1) and applied (\ref{toto.e22}) to $B_{i}$,$g_{i}$, $\dot{B_{i}}$ and $\dot{g_{i}}$.  This proves (1) by the Cook argument.

Let now $B\in\mcL_{0}$, satisfying (\ref{transfer-to-hyperboloid}), and $\Delta= - \supp(\widehat B)\cap \Sp \,U\subset H_{m}$.
We fix $O\subset \rr^{1+d}$, which is an arbitrarily small neighborhood of $\Delta$, and a function $h\in\cS(\rr^{1+d})$ with $\supp \widehat{h}\subset O$ and $\widehat{h}= (2\pi)^{-(d+1)/2}$ on $\Delta$. Setting $C^{*}= \int B^{*}(t,x)h(t,x)dtdx$ we have: $C\in \mcL_{0}$ and:
\[
 \widehat{C^{*}}(E, p)= (2\pi)^{(d+1)/2} \widehat{h}(E,p)\widehat{B^{*}}(E,p), \ C^{*}\Omega= (2\pi)^{(d+1)/2} \widehat{h}(H, P)B^{*}\Omega.
\]
This implies that $-\supp(\widehat C)\subset O$, and
\beq\label{toto.e23}
\begin{array}{rl}
 B^{*}_{t}(g_t)\Omega&= (2\pi)^{d/2}\widehat{f}(P)B^{*}\Om= (2\pi)^{d/2}\widehat{f}(P)\one_{\Delta}(U)B^{*}\Om\\[2mm]
&=(2\pi)^{d/2}\widehat{f}(P)(2\pi)^{(d+1)/2}\widehat{h}(H, P)B^{*}\Om= (2\pi)^{d/2}\widehat{f}(P)C^{*}\Om= C^{*}_{t}(g_t)\Omega.
\end{array}
\eeq
Introducing observables $C_{i}$ as above for $B_{i}$ and using also (\ref{toto.e22}) and Lemma \ref{toto.21} (2) we obtain that
\begin{equation}
\label{toto.e24}
\Psi^{+}= \lim_{t\to\infty} B^{*}_{1,t}(g_{1,t})B^{*}_{2,t}(g_{2,t})\Om=\lim_{t\to\infty} C^{*}_{1,t}(g_{1,t})C^{*}_{2,t}(g_{2,t})\Om.
\end{equation}
Thus we can assume that the energy-momentum  transfers of $B_{i}^{*}$ entering in the construction of scattering states are localized in arbitrarily small neighborhoods of subsets of $H_{m}$. This observation will be important in the proof of (2) to which we now proceed.

Let $\tilde \Psi_t=\tilde{B}^{*}_{1,t}(\tilde g_{1,t})\tilde{B}^{*}_{2,t}(\tilde g_{2,t})\Om$ be the approximants of the scattering state $\tilde \Psi^+$. In order
to compute the scalar product $( \tilde \Psi_t|\Psi_t)$ we first observe that
\beqa
[[\tilde{B}_{1,t}(\tilde g_{1,t}),B^{*}_{1,t}(g_{1,t})],B^{*}_{2,t}(g_{2,t}) ]\in O(t^{-\infty}). \label{double-commutator}
\eeqa
This relation can be justified by writing $\tilde g_1=\tilde g_{1,1}+\tilde g_{1,2}$, where $\tilde g_{1,i}$ are positive energy KG solutions 
such that  the velocity support of $\tilde g_{1,i}$ and $g_i$ are disjoint for $i=1,2$.  Then (\ref{double-commutator})
follows from (\ref{toto.e22}) and the Jacobi identity. Next we note that
\beqa
\tilde  B_{i,t}(\tilde g_{i,t}) B^{*}_{j,t}(g_{j,t})\Om=\Om( \Om| \tilde B_{i,t}(\tilde g_{i,t})  B^{*}_{j,t}(g_{j,t})\Om) \label{factorization}, \ 1\leq  i,j\leq 2.
\eeqa
This relation follows from the fact that  $\tilde  B_{i,t}(\tilde g_{i,t})  B^{*}_{j,t}(g_{j,t})\Om$ belongs 
to the range of $\one_{-K_{j}+\tilde{K}_{i}}(U)$, where $K_j$ and $\tilde K_i$ are the energy-momentum transfers of $B_j$ and $\tilde{B}_i$,
respectively. In view of  (\ref{toto.e24}) $-K_{j}$, $-\tilde{K}_{i}$ can be chosen in arbitrarily small neighbourhoods of $H_m$.
 Since a non-zero vector which is a difference of two vectors from $H_m$ is space-like, (\ref{factorization}) follows.

\def\tB{{\tilde{B}}}
We set for simplicity of notation $B_{i}(t):= B_{i,t}(g_{i,t})$, $\tilde{B}_{j}(t):= \tilde{B}_{j,t}(\tilde{g}_{j,t} )$. Then
\begin{equation}
\label{toto.e25}
\begin{array}{rl}
(\tilde{\Psi}_{t}| \Psi_{t})=&(\Om| \tB_{2}(t)B_{1}^{*}(t)\tB_{1}(t)B_{2}^{*}(t)\Om)\\[2mm]
&+(\Omega| \tB_{2}(t)B_{2}^{*}(t)\tB_{1}(t)B_{1}^{*}(t)\Omega)\\[2mm]
&+(\Omega\ \tB_{2}(t)[[\tB_{1}(t),B_{1}^{*}(t)], B_{2}^{*}(t)]\Om).
\end{array}
\end{equation}
Making use of (\ref{double-commutator}) and (\ref{factorization}), we conclude the proof of (\ref{scalar-product}).  It
follows immediately from~(\ref{scalar-product}) that the scattering states $\Psi^+$ depend only on the single-particle states $\Psi_i$
(and not on  a particular choice of $B_i$ and $g_i$). Finally, 
relation~(\ref{energy-factorization-relation}) is an easy consequence of Lemma \ref{toto.21} (1). \qed

\section{Proof of Theorem \ref{Weak-asymptotic-completeness}}\init\label{blit}
In the next proposition we will use the notation $N_{B}(h, t)$ introduced in (\ref{def-de-Nb}) for $B\in \mcL_{0}$ and $h\in \coinf(\rr^{d})$.

\begin{proposition}\label{C1}
 Let  $i= 1,2$, $\Delta_{i}\Subset H_{m}$  with $\Delta_{1}, \Delta_{2}$ disjoint and $B_{i}\in \mcL_{0}$ with 
$\supp(\widehat B_{1})$,  $\supp(\widehat B_{2})$ disjoint. Assume moreover that:
 \begin{eqnarray}
\label{e.c1a}
-\supp(\widehat B_{i})\cap \Sp\, U\subset \Delta_{i},\\[2mm]
\label{e.c1b}(\Delta_{i}+ \supp(\widehat B_{i}))\cap \Sp (U)\subset \{0\},  \ i=1,2,\\[2mm]
\label{e.c1c}(\Delta_{i}+ \supp(\widehat B_{j}))\cap \Sp (U)= \emptyset, \ i\neq j.
\end{eqnarray}
 Let $h_{i}\in \coinf(\rr^{d})$ with disjoint supports and $h_{i}\equiv 1$ on ${\rm Vel}(\Delta_{i})$. Then for $\Psi_{i}\in \one_{\Delta_{i}}(U)\cH$ one has:
\begin{equation}
\label{e.c4}
\lim_{t\to +\infty} N_{B_{1}}(h_{1}, t)N_{B_{2}}(h_{2}, t)W_{2}^{+}(\Psi_{1}\otimes_{\rm s}\Psi_{2})= W_{2}^{+}(N_{B_{1}}(\one)\Psi_{1}\otimes_{\rm s}N_{B_{2}}(\one)\Psi_{2}).
\end{equation}
\end{proposition}
\begin{remark}
  Note that  $W_{2}^{+}(\Psi_{1}\otimes_{\rm s}\Psi_{2})$ belongs to $\cH_{\rm c}(U)$, and that $N_{B_{i}}(\one)\Psi_{i}$ belong to $\one_{\Delta_{i}}(U)\cH$, because of (\ref{e.c1a}), (\ref{e.c1b}),  hence  all the expressions appearing in (\ref{e.c4}) are well defined.
 \end{remark}
 
\proof
 We first claim that for $B,\Delta, \Psi, h$  as in the proposition one has:
 \begin{equation}
\label{e.c0}
\lim_{t\to+\infty}N_{B}(h, t)\Psi= N_{B}(\one)\Psi.
\end{equation}
In fact we first note that because of (\ref{e.c1a}), (\ref{e.c1b}) we have 
\beq\label{tarali}
B^{*}B\one_{\Delta}(U)=B^{*}|\Omega\rangle\langle \Omega|B\one_{\Delta}(U) =\one_{\Delta}(U)B^{*}B \one_{\Delta}(U).
\eeq Therefore
\[
\begin{array}{rl}
N_{B}(h, t)\Psi=& \e^{\i tH}N_{B}(h_{t})\e^{-\i tH}\Psi\\[2mm]
=&\e^{\i t\omega(P)} a_{B}^{*}\circ (\one_{\cH}\otimes h_{t})\circ a_{B}\e^{- \i t \omega(P)}\Psi\\[2mm]
=& a_{B}^{*}\circ \e^{\i t \omega(P+ D_{x})}(\one_{\cH}\otimes h_{t})\e^{-\i t \omega(P+D_{x})} \circ a_{B}\Psi, 
\end{array}
\]
using (\ref{toto.e0}). Since $ \e^{\i t \omega(P+ D_{x})}x \e^{-\i t \omega(P+ D_{x})}= x+ t\nabla \omega(P+ D_{x})$, we have
\[
 \e^{\i t \omega(P+ D_{x})}(\one_{\cH}\otimes h_{t})\e^{-\i t \omega(P+D_{x})}= h\left(\frac{x}{t}+ \nabla \omega(P+ D_{x})\right),
\]
from which we easily deduce that
\[
\slim_{t\to +\infty}  \e^{\i t \omega(P+ D_{x})}(\one_{\cH}\otimes h_{t})\e^{-\i t \omega(P+D_{x})}= h(\nabla \omega(P+ D_{x})).
\]
Inserting as usual energy-momentum projections, this implies that
\[
\lim_{t\to +\infty}N_{B}(h, t)\Psi=a_{B}^{*}\circ h(\nabla \omega(P+ D_{x}))\circ a_{B}\Psi= a_{B}^{*}a_{B} h(\nabla \omega(P))\Psi,
\]
using once again (\ref{toto.e0}). From the support property of $h$ we have $h(\nabla \omega(p))=1$ for $(\omega(p), p)\in \Delta$, hence
$h(\nabla \omega(P))\Psi= \Psi$, which completes the proof of (\ref{e.c0}).

We now proceed to the proof of (\ref{e.c4}).  Since $N_{B_{1}}(h_{1}, t)N_{B_{2}}(h_{2}, t)\one_{\Delta_{1}}(U)$ is uniformly bounded in time for any $\Delta_{1}\Subset \rr^{1+d}$, it suffices by density to assume that $\Psi_{i}= A^{*}_{i, t}(g_{i,t})\Omega$ for $A_{i}\in \mcL_{0}$ satisfying (\ref{transfer-to-hyperboloid}) and $g_{i}$ a positive energy KG solution with the velocity support 
included in ${\rm Vel}(\Delta_{i})$, so that $\Psi_{i}= \one_{\Delta_{i}}(U)\Psi_{i}$. Let us fix such $A_{i}, g_{i}$.

By (\ref{e.c1c}) we have 
$B_{i}A_{j}^{*}\Omega=0$ if $i\neq j$, hence:
\begin{equation}
\label{e.c9}
N_{B_{i}}(h_{i}, t)A^{*}_{j, t}(g_{j,t})\Omega=0, \ i\neq j.
\end{equation}
Next we note that for $i\neq j$:
\begin{equation}
\label{e.c10}
\| [N_{B_{i}}(h_{i}, t), A_{j,t}^{*}(g_{j,t})]\| \in O(t^{-\infty}).
\end{equation}
In fact since  the support of $h_{i}$ and the velocity support of $g_{j}$ are disjoint, we can pick  a smooth partition of unity $1= \chi_{i}(x)+ \chi_{j}(x)$ with $\chi_{i}\equiv 0$ near the velocity support of $g_{j}$ and $\chi_{j}\equiv 0$ near  the support of $h_{i}$. We have then
by almost locality
\[
\begin{array}{rl}
\|  [N_{B_{i}}(h_{i}, t), A_{j,t}^{*}(g_{j,t})]\|\leq& \int \| [(B_{i}^{*}B_{i})(t,x), A_{j}^{*}(t,y)]\|  |h_{i}(\xt)| | g_{j}(t, y)| dx dy\\[2mm]
&\leq C_{N}\int \langle x- y\rangle^{-N}| h_{i}(\xt)| | g_{j}(t,y)|\chi_{j}(\frac{y}{t})dxdy\\[2mm]
&+ C_{N}\int\langle x-y\rangle^{-N}|h_{i}(\xt)| |g_{j}(t,y)|\chi_{i}(\frac{y}{t})dxdy. 
\end{array}
\]
The first integral is $O(t^{-\infty})$ because $h_{i}$ and $\chi_{j}$ have disjoint supports, the second is also $O(t^{-\infty})$ using that $\supp \chi_{i}$ is disjoint from the velocity support of $g_{j}$ and applying Prop. \ref{toto.20} (2). This proves (\ref{e.c10}). 

Finally since $N_{B_{i}}(\one)\Psi_{i}\in\one_{\Delta_{i}}(U)\cH$, we can find for any $0<\epsilon_{i}\ll 1$ operators $\tilde{A}_{i}\in \mcL_{0}$ and positive energy solutions $\tilde{g}_{i}$ satisfying  the same properties as $A_{i}, g_{i}$ such that  
\begin{equation}
\label{e.c11}
\| N_{B_{i}}(\one)\Psi_{i}- \tilde{A}_{i,t}^{*}(\tilde{g}_{i,t})\Omega\|\leq \epsilon_{i}, \  i=1,2.
\end{equation}
Using successively (\ref{e.c10}), (\ref{e.c0}) and (\ref{e.c11}), we obtain:
\[
\begin{array}{rl}
N_{B_{1}}(h_{1}, t)N_{B_{2}}(h_{2}, t)(\Psi_{1}\timeso\Psi_{2})= &N_{B_{1}}(h_{1}, t)N_{B_{2}}(h_{2}, t)A_{1,t}^{*}(g_{1,t})A_{2, t}^{*}(g_{2,t})\Omega+ o(t^{0})\\[2mm]
=& N_{B_{1}}(h_{1}, t)A_{1,t}^{*}(g_{1,t})N_{B_{2}}(h_{2}, t)A_{2, t}^{*}(g_{2,t})\Omega+ o(t^{0})\\[2mm]
=&N_{B_{1}}(h_{1}, t)A_{1,t}^{*}(g_{1,t})N_{B_{2}}(\one)\Psi_{2}+ o(t^{0})\\[2mm]
=& N_{B_{1}}(h_{1}, t)A_{1,t}^{*}(g_{1,t})\tilde{A}_{2,t}^{*}(\tilde{g}_{2,t})\Omega + o(t^{0})+ O(t^{0})\epsilon_{2}.
\end{array}
\]
Using then (\ref{toto.e22}), (\ref{e.c10}), (\ref{e.c0}),  we have:
\[
\begin{array}{rl}
&N_{B_{1}}(h_{1}, t)A_{1,t}^{*}(g_{1,t})\tilde{A}_{2,t}^{*}(\tilde{g}_{2,t})\Omega= N_{B_{1}}(h_{1}, t)\tilde{A}_{2,t}^{*}(\tilde{g}_{2,t})A_{1,t}^{*}(g_{1,t})\Omega+ o_{\epsilon_{2}}(t^{0})\\[2mm]
=&  \tilde{A}_{2,t}^{*}(\tilde{g}_{2,t})N_{B_{1}}(h_{1}, t)A_{1,t}^{*}(g_{1,t})\Omega+ o_{\epsilon_{2}}(t^{0})
=  \tilde{A}_{2,t}^{*}(\tilde{g}_{2,t})N_{B_{1}}(\one)\Psi_{1}+ o_{\epsilon_{2}}(t^{0})\\[2mm]
=& \tilde{A}_{2,t}^{*}(\tilde{g}_{2,t}) \tilde{A}_{1, t}^{*}(\tilde{g}_{1,t})\Omega + o_{\epsilon_{2}}(t^{0})+ O_{\epsilon_{2}}(t^{0})\epsilon_{1}
= \tilde{A}_{1, t}^{*}(\tilde{g}_{1,t})\tilde{A}_{2,t}^{*}(\tilde{g}_{2,t})\Omega + o_{\epsilon_1,\epsilon_{2}}(t^{0})+ O_{\epsilon_{2}}(t^{0})\epsilon_{1}\\[2mm]
=& \tilde{\Psi}_{1}\timeso\tilde{\Psi}_{2}+ o_{\epsilon_{1}, \epsilon_{2}}(t^{0})+ O_{\epsilon_{2}}(t^{0})\epsilon_{1},
\end{array}
\]
for $\tilde{\Psi}_{i}=  \tilde{A}_{i,t}^{*}(\tilde{g}_{i,t})\Omega$. By Prop. \ref{def-de-wave} (1) we have also
\[
\| N_{B_{1}}(\one)\Psi_{1}\timeso N_{B_{2}}(\one)\Psi_{2}- \tilde{\Psi}_{1}\timeso\tilde{\Psi}_{2}\| \leq C (\epsilon_{1}+ \epsilon_{2}).
\]
We obtain finally
\[
\begin{array}{rl}
&N_{B_{1}}(h_{1}, t)N_{B_{2}}(h_{2}, t)(\Psi_{1}\timeso\Psi_{2})\\[2mm]
=&N_{B_{1}}(\one)\Psi_{1}\timeso N_{B_{2}}(\one)\Psi_{2}+ o_{\epsilon_{1}, \epsilon_{2}}(t^{0})+ O(\epsilon_{1}+ \epsilon_{2}) + O_{\epsilon_{2}}(t^{0})\epsilon_{1}.
\end{array}
\]
Picking first $\epsilon_{2}\ll 1$, then $\epsilon_{1}\ll 1$ and then $t\gg 1$, we obtain (\ref{e.c4}). \qed
 
\medskip

\begin{lemma}
 \label{Delta-inclusions}
  Let $\Delta\subset G_{2m}$ be an open bounded set. Then
\[
\one_{\Delta}(U)\cH_{2}^{+}= {\rm Span}\{W_{2}^{+}(\Psi_{1}\otimes_{\rm s}\Psi_{2})\ : \ \Psi_{i}\in \one_{\Delta_{i}}(U)\cH, \ \Delta_{i}\Subset H_{m}, \ \Delta_{1}+ \Delta_{2}\subset \Delta, \ \Delta_{1}\cap \Delta_{2}= \emptyset\}^{\rm cl}.
\]
\end{lemma}
\proof The proof follows immediately from Prop. \ref{def-de-wave} (2) and the absolute continuity of the spectral measure of $(H,P)$ restricted to $\cH_{m}$ recalled in its proof. \qed

\medskip

\bel\label{O-lemma} Let $\De\subset G_{2m}$ be an open bounded set s.t. $(\ov{\De}-\ov{\De})\cap\Sp\,U= \{0\}$. Let $\De_1,\De_2\Subset H_{m}$ be  disjoint and such that  $\De_1+\De_2\subset \De$.  Then there exist 
$O_1, O_2\subset\real^{1+d}$ which are disjoint open neighbourhoods of $\De_1,\De_2$, respectively, such that for any 
 $K_1, K_2\Subset\real^{1+d}$ satisfying  $-K_i\subset O_i$, $-K_i\cap \Sp\, U\subset \De_i$, one has:
\beqa
& &(\ov{\De}+K_1+K_2)\cap \Sp\, U\subset \{0\}, \label{transfer-to-vacuum-one}\\
& &-(K_1+K_2)\subset \De, \label{transfer-from-vacuum-one}\\
& &(\De_i+K_i)\cap\Sp\,U\subset \{0\}, \label{smallness-of-transfers-one}\\
& &(\De_i+K_j)\cap \Sp\, U=\emptyset, \quad i\neq j. \label{mixed-terms-vanishing-one}
\eeqa

\eel
\proof Assume that  $O_i\subset \De_i+B(0,\eps)$, where $B(0,\eps)$ is the ball of radius $\eps$ centered at zero.
To prove (\ref{transfer-to-vacuum-one}), we write
\beqa
\ov{\De}+K_1+K_2\subset \ov{\De}-O_1-O_2&\subset& \ov{\De}-\De_1-\De_2+B(0,2\eps)\non\\
&\subset&  \ov{\De}-\ov{\De}+B(0,2\eps).
\eeqa 
Since, by assumption,  $(\ov{\De}-\ov{\De})\cap\Sp\, U= \{0\}$  and $0$ is isolated in $\Sp\, U$, 
we obtain that $(\ov{\De}-\ov{\De}+B(0,2\eps))\cap\Sp\,U=\{0\}$ for  $\eps\ll 1$.
As for (\ref{transfer-from-vacuum-one}), we obtain that 
\beqa
-(K_1+K_2)\subset O_1+O_2\subset \De_1+\De_2+B(0,2\eps)\subset \De,
\eeqa 
for $\eps\ll 1$  using that $\Delta_{i}$ are compact and $\De$ is open.
Finally we write:
\beqa
\De_i+K_j\subset O_i-O_j\subset \Delta_{i}- \Delta_{j}+ B(0, 2\epsilon).
\eeqa 
We note  that  a difference of two vectors from $H_m$ is  either $0$  or space-like.  For $\eps\ll 1$ we obtain (\ref{smallness-of-transfers-one}) if $i=j$ and (\ref{mixed-terms-vanishing-one}) if $i\neq j$.\qed

\medskip

\bel\label{subspace-equality-lemma} Let $\De\Subset H_m$  and $O\subset \real^{1+d}$ be a sufficiently small neighbourhood 
of $\De$. Then 
\[
\one_{\Delta}(U)\hil=\Span\{\, N_{B}(\one)\one_{\Delta}(U)\hil   \setbar B\in \mcL_0,\,  -\supp(\widehat B)\subset O, 
-\supp(\widehat B)\cap\Sp\,U\subset \De\, \}^{\cl}.
\]
\eel
\proof  Arguing as in the proof of (\ref{smallness-of-transfers-one}) we  fix $O$ sufficiently small such that for all $B$  in the lemma one has $(\Delta+\supp(\widehat B))\cap \Sp\, U= \{0\}$. 
  Let  now $S$ be the subspace in the r.h.s. of the equality stated in the lemma and let $P_S$ be the corresponding projection.  By (\ref{tarali}) we have   $P_{S}\leq \one_{\Delta}(U)$. 
 To complete the proof  we adapt an argument from the proof of  \cite[Thm.~3.5]{DT11}. Assume that $P_{S}\neq \one_{\Delta}(U)$ and let   $\Psi\neq 0$ with $\Psi= \one_{\Delta}(U)\Psi$, $P_{S}\Psi=0$.  Clearly  there exists $f\in \cS(\rr^{1+d})$ such that $\supp \widehat{f}\subset -O$ and $\widehat{f}(-H, -P)\Psi\neq 0$. By cyclicity of the vacuum there exists $A\in\mfa(\mco)$, for some open bounded $\mco\subset\real^{1+d}$, such that:
 \beqa
0\neq ( A^{*}\Om| \widehat{f}(-H,-P)\Psi)=(\Om|B\Psi), \hbox{ for }B:=(2\pi)^{-\fr{1+d}{2}}\int f(t,x)A(t,x)dtdx.
\eeqa  
 Since $\widehat{B}(E, p)= \widehat{f}(E, p)\widehat{A}(E, p)$ we see that $B$ satisfies the conditions from the lemma, and $B\Psi\neq 0$. By the norm continuity of $x\mapsto B(x)$ this implies that 
$(\Psi| N_{B}(\one)\Psi)\neq 0$  which contradicts the fact that $P_{S}\Psi=0$. \qed

\medskip

\noindent {\em Proof of Thm. \ref{Weak-asymptotic-completeness}. } In view of Thm.~\ref{Main-theorem}, it suffices to verify the inclusion
\beqa
\one_{\Delta}(U)\cH_{2}^{+}\subset \Span\{  
\Ran\, \C_{2,\al}^+(\De)\setbar     \al\in J  \}^{\cl}.
\eeqa
By  Lemma~\ref{Delta-inclusions}, it is enough to show that for any  $\De_1, \De_2\Subset H_m$  such that 
$\De_1+\De_2\subset \De$ and $\De_1\cap \De_2=\emptyset$  one has
\beqa
W_{2}^{+}\left(\one_{\Delta_{1}}(U)\cH\otimes_{\rm s}\one_{\Delta_{2}}(U)\cH\right)\subset  \Span\{  \Ran\, \C_{2,\al}^+(\De)\setbar     \al\in J  \}^{\cl}. \label{small-inclusion}
\eeqa 
Let $O_1,O_2\in \real^{1+d}$ be sufficiently small open neighbourhoods of $\De_1,\De_2$, respectively,
so that the assertions of Lemma~\ref{O-lemma} hold. We choose $B_1, B_2\in \mcL_0$, such that   $-\supp(\widehat B_{i})\subset O_i$, $-\supp(\widehat B_{i})\cap \Sp\, U\subset \De_i$.
By Lemma~\ref{O-lemma}, $B_1, B_2$ are $\De-$admissible in the sense of Definition~\ref{delta-admissible} and satisfy the assumptions of Prop. \ref{C1}. 
Finally, we choose $h_1,h_2\in C_0^{\infty}(\real^d)$ 	as in  Prop. \ref{C1}. 

Let $J_0$ be the set of quadruples $(B_1,B_2, h_1, h_2)$ as specified above.   We get
\beqa
&&\Span \{\C_{2,\alpha}^+(\De)\circ W_{2}^{+}\left(\one_{\Delta_{1}}(U)\cH\otimes_{\rm s} \one_{\Delta_{2}}(U)\cH\right)\ : \ \alpha\in J_{0} \}
\nonumber\\
 &=&\Span\{W_{2}^{+}\left(N_{B_{1}}(\one)\one_{\Delta_{1}}(U)\cH\otimes_{\rm s}N_{B_{2}}(\one)\one_{\Delta_{2}}(U)\cH\right)\ : \ \alpha\in J_{0}\}\label{subspace-inclusion}\\
 &=& W_{2}^{+}\left(\one_{\Delta_{1}}(U)\cH\otimes_{\rm s}\one_{\Delta_{2}}(U)\cH\right).\nonumber
\eeqa
In the first step we  use  Prop. \ref{C1} and in the second  Lemma~\ref{subspace-equality-lemma}.
Clearly, $J_0\subset J$, thus the subspace on the l.h.s. of (\ref{subspace-inclusion}) is included in the subspace on the r.h.s. of (\ref{small-inclusion}).
This concludes the proof. \qed

\appendix

\section{Propagation estimates for inhomogeneous evolution equations}\init\label{alo}
In this section we extend standard results on propagation estimates and existence of limits for unitary propagators to the case of an inhomogeneous evolution equation:
\[
\p_{t}u(t)= -\i H u(t)+ r(t).
\]
Let $\cH$ be a Hilbert space and $H$ a self-adjoint operator on $\cH$.  We fix a function 
\[
\rr^{+}\ni t\mapsto u(t)\in \cH,
\]
 such that
\begin{equation}
\label{abs.e1}
\begin{array}{rl}
i)& \sup_{t\geq 0}\| u(t)\|<\infty,\\[2mm]
ii)& u(t)\in C^{1}(\rr^{+}, \cH)\cap C^{0}(\rr^{+}, \Dom H),
\end{array}
\end{equation}
and set:
\[
r(t):= \p_{t}u(t)+ \i  H u(t).
\]
For a map  $\rr^{+}\ni t\mapsto M(t)\in B(\cH)$ we denote by $\D M(t)= \p_{t}M(t)+ [H, \i M(t)]$ the Heisenberg derivative of $M(t)$, w.r.t. the evolution $\e^{-\i tH}$. We assume that  $[H, \i M(t)]$, defined first as a quadratic form on $\Dom H$, extends by continuity to a bounded operator.

The following three lemmas can be proved by mimicking standard arguments, see e.g. \cite[Sect. B.4]{DG97}.
By $C_j(\,\cdot\,)$, $B(\,\cdot\,)$, $B_1(\,\cdot\,)$ we denote auxiliary functions from $\real^+$ to $B(\hil)$.
\begin{lemma}\label{A1}
 Let $\rr^{+}\ni t\mapsto M(t)\in B(\cH)$ be such that:
 \[
\begin{array}{rl}
i)&\sup_{t\in \rr^{+}}\|M(t)\|<\infty,\  \|M(\cdot)r(\cdot)\|, \ \|M^{*}(\cdot)r(\cdot)\|\in L^{1}(\rr^{+},dt),\\[2mm]
ii)&\D M(t)\geq B^{*}(t)B(t)- \sum_{j=1}^{n}C_{j}^{*}(t)C_{j}(t),\ \int_{\rr^{+}}\| C_{j}(t)u(t)\|^{2}dt<\infty.
\end{array}
\]
Then 
\[
 \int_{0}^{+\infty} \| B(t)u(t)\|^{2}dt<\infty.
\]
\end{lemma}
\begin{lemma}\label{A3}
 Let $\rr^{+}\ni t\mapsto M(t)\in B(\cH)$ be such that:
 \[
\begin{array}{rl}
i)&\sup_{t\in \rr^{+}}\|M(t)\|<\infty,\  \|M(\cdot)r(\cdot)\|, \ \|M^{*}(\cdot)r(\cdot)\|\in L^{1}(\rr^{+},dt),\\[2mm]
ii)&|(u_{1}|\D M(t)u_{2})|\leq\sum_{j=1}^{n}\|C_{j}(t)u_{1}\|\|C_{j}(t)u_{2}\|,\ u_{1}, u_{2}\in\cH, \\[2mm]
\hbox{ with }&\int_{\rr^{+}}\| C_{j}(t)u(t)\|^{2}dt<\infty.
\end{array}
\]
Then 
\[
\lim_{t\to +\infty}(u(t)| M(t)u(t))\hbox{ exists}.
\]
\end{lemma}
\begin{lemma}\label{A2}
Let $\rr^{+}\ni t\mapsto M(t)\in B(\cH)$ be such that:
\[
\begin{array}{rl}
i)&\| M(\cdot)r(\cdot)\|\in L^{1}(\rr^{+},dt),\\[2mm]
ii)&|(u_{1}| \D M(t)u(t))|\leq \| B_{1}(t)u_{1}\|\| B(t)u(t)\|,\hbox{ with}\\[2mm]
iii)&\int_{\rr^{+}} \| B(t)u(t)\|^{2}dt <\infty, \ \int_{\rr^{+}}\| B_{1}(t)\e^{-\i tH}u_{1}\|^{2}dt\leq C \| u_{1}\|^{2}, \ u_{1}\in \cH.
\end{array}
\]
Then 
\[
\lim_{t\to +\infty}\e^{\i tH}M(t)u(t)\hbox{ exists}.
\]
\end{lemma}



\begin{thebibliography}{LNT2}
\bibitem[AH67]{AH67} H. Araki and R. Haag:
\emph{Collision cross sections in terms of local observables}.
Commun. Math. Phys. \bf 4\rm, (1967) 77--91.
\bibitem[Ar99]{Ar99} H. Araki: \emph{Mathematical theory of quantum fields}.
Oxford Science Publications, 1999.
\bibitem[Ar74]{Ar74} W. Arveson: \emph{On groups of automorphisms of operator algebras}.
J.  Funct. Anal. \bf 15\rm, (1974) 217--243.
\bibitem[Ar82]{Ar82} W. Arveson:
\emph{The harmonic analysis of automorphism groups}. In
Operator algebras and applications, Part~I (Kingston, Ont., 1980),
Proc. Sympos. Pure Math., 38,  Amer. Math. Soc., Providence, R.I.,1982.D., pp. 199--269.
\bibitem[Bu90]{Bu90} D. Buchholz:
\emph{Harmonic analysis of local operators}.
Commun. Math. Phys. \bf 129\rm, (1990) 631--641.
\bibitem[BF82]{BF82} D. Buchholz and K. Fredenhagen: \emph{Locality and the structure of particle states}. 
Commun. Math. Phys. \bf 84\rm, (1982) 1--54.
\bibitem[BPS91]{BPS91} D. Buchholz, M. Porrmann and U. Stein:
\emph{Dirac versus Wigner: Towards a universal particle
concept in quantum field theory}.
Phys. Lett. B  \bf  267\rm, (1991) 377--381.
\bibitem[CD82]{CD82}   M. Combescure and F. Dunlop: \emph{Three-body asymptotic completeness for $P(\phi)_2$ models}. 
Commun. Math. Phys. \bf 85\rm, (1982) 381--418.
\bibitem[Dy05]{Dy05} W. Dybalski: \emph{Haag-Ruelle scattering theory in presence of
massless particles}.  Lett. Math. Phys. \bf 72\rm, (2005) 27--38.
\bibitem[Dy10]{Dy10} W. Dybalski: \emph{Continuous spectrum of automorphism groups and the infraparticle problem}.
Commun. Math. Phys. \bf 300\rm, (2010) 273--299. 
\bibitem[DM12]{DM12} W. Dybalski and J.S. M\o ller: \emph{The translation invariant massive Nelson model: III. Asymptotic completeness below the two-boson threshold.} Preprint arXiv:1210.6645 [math-ph].
\bibitem[DT11a]{DT11} W. Dybalski and Y. Tanimoto:  
\emph{Asymptotic completeness for infraparticles in two-dimensional conformal field theory.}  Preprint arXiv1112.4102 [math-ph].
\bibitem[DT11b]{DT11.1} W. Dybalski and Y. Tanimoto:  
\emph{Infraparticles with superselected direction of motion in two-dimensional conformal field theory.}
Commun.  Math. Phys. \bf 311\rm, (2012) 457--490.
\bibitem[De93]{De93} J. Derezi\'nski:
\emph{Asymptotic completeness of long-range $N$-body quantum systems}.
Ann. of Math. \bf 138\rm, (1993) 427--476.
\bibitem[DG99]{DG99}  J.~Derezi{\'n}ski and C.~G{\'e}rard:
 \emph{Asymptotic completeness in quantum  field theory. Massive {Pauli-Fierz} {Hamiltonians}}. Rev. Math. Phys.
  \textbf{11}, (1999) 383--450.
\bibitem[DG97]{DG97}   J.~Derezi{\'n}ski and C.~G{\'e}rard:  \emph{Scattering theory of classical and quantum $N$-particle systems}.
    Springer, 1997.
\bibitem[DG00]{DG00} J. Derezi\'nski and C. G\'erard: \emph{Spectral and scattering theory of spatially cut-off $P(\phi)_2$
Hamiltonians}. Commun. Math. Phys. \bf 213\rm, (2000) 39--125.
\bibitem[En75]{En75} V. Enss: \emph{Characterization of particles by means of local observables}. Commun.
Math. Phys. \bf 45\rm, (1975) 35--52.
\bibitem[En78]{En78} V. Enss:
\emph{Asymptotic completeness for quantum mechanical
potential scattering}.
Commun. Math. Phys. \bf 61\rm, (1978) 285--291.
\bibitem[FGS02]{FGS02} J.~Fr{\"o}hlich, M.~Griesemer and B.~Schlein:
  \emph{Asymptotic completeness for {R}ayleigh scattering}.
  Ann. Henri Poincar{\'e} \textbf{3}, (2002) 107--170.
\bibitem[FGS04]{FGS04} J.~Fr{\"o}hlich, M.~Griesemer and B.~Schlein:
  \emph{Asymptotic completeness for {C}ompton scattering}.
  Commun. Math. Phys. \textbf{252}, (2004) 415--476.
\bibitem[GJS73]{GJS73}   J. Glimm,  A. Jaffe  and  T. Spencer: \emph{The particle structure of the weakly coupled $P(\phi)_2$
model and other applications of high temperature expansions: Part I. Physics of quantum field models. Part II. The cluster expansion.} 
In: Constructive quantum field theory. (Erice, 1973), G. Velo, A. S. Wightman (eds.). Berlin, Heidelberg, New York: Springer 1973.
\bibitem[Ge91]{Ge91} C. G\'erard: \emph{Mourre estimate for regular dispersive systems}, Ann. Inst. H. Poincar\'e \textbf{54}, (1991) 59–-88.
\bibitem[Gr90]{Gr90} G. M. Graf:
\emph{Asymptotic completeness for N-body short-range 
quantum systems: a new proof}.
Commun. Math. Phys. \bf 132\rm, (1990) 73--101.
\bibitem[Ha58]{Ha58} R. Haag: \emph{Quantum field theories with composite particles
and asymptotic conditions}.  Phys. Rev. \bf 112\rm, (1958) 669--673.
\bibitem[Ha]{Ha} R. Haag:
\emph{Local quantum physics}. Springer, 1992.
\bibitem[Po04a]{Po04.1} M. Porrmann:
\emph{Particle weights and their disintegration I}.
Commun. Math. Phys. \bf 248\rm, (2004) 269--304.
\bibitem[Po04b]{Po04.2} M. Porrmann:
\emph{Particle weights and their disintegration II}.
Commun. Math. Phys. \bf 248\rm, (2004) 305--333.
\bibitem[Ru62]{Ru62} D. Ruelle:  \emph{On the asymptotic condition in quantum 
field theory}. Helv. Phys. Acta \bf 35\rm, (1962) 147--163.
\bibitem[RS3]{RS3} M. Reed and B. Simon:
\emph{Methods of modern mathematical physics.
Part III: Scattering theory.}
Academic Press, 1979.
\bibitem[SiSo87]{SiSo87} I. M. Sigal and A. Soffer:
\emph{The N-particle scattering problem: asymptotic completeness for short-range systems}.
Ann. of Math. \bf 126\rm, (1987) 35--108.
\bibitem[SZ76]{SZ76} T. Spencer and F. Zirilli: \emph{Scattering states and bound states in $\la P(\phi)_2$}. Commun. Math. Phys. \bf 49\rm, (1976) 1--16.
\bibitem[Zi97]{Zi97} L. Zieli\'nski: \emph{Scattering for a dispersive charge-transfer model.}
Ann. Inst. Henri Poincar\'e \bf  67\rm, (1997)  339--386. 
\end{thebibliography}
\end{document}